\documentclass[hidelinks,lettersize,journal,final]{IEEEtran}

\makeatletter
\newcommand\ifACM[1]{\@ifclassloaded{acmart}{#1}{}}
\newcommand\ifIEEE[1]{\@ifclassloaded{IEEEtran}{#1}{}}
\makeatother

\ifACM{%
\setcopyright{acmcopyright}
\copyrightyear{2018}
\acmYear{2018}
\acmDOI{XXXXXXX.XXXXXXX}

\acmJournal{JACM}
\acmVolume{37}
\acmNumber{4}
\acmArticle{111}
\acmMonth{8}

\newcommand\IEEEPARstart[2]{#1#2\xspace}
}
\ifIEEE{}

\usepackage[english]{babel}

\usepackage[utf8]{inputenc}
\usepackage[T1]{fontenc}

\usepackage{verbatim}

\ifIEEE{\usepackage{orcidlink}}

\PassOptionsToPackage{hyphens}{url}

\usepackage{url}                %
\usepackage{graphicx}           %

\usepackage[
    binary-units,
    range-units=single,
    list-units=single,
    per-mode=symbol,
    per-symbol=p,
    detect-all,
]{siunitx}                      %
\usepackage{relsize}            %
\usepackage{xspace}             %
\usepackage{calc}               %
\usepackage{microtype}          %
\usepackage[inline]{enumitem}   %
\usepackage{nth}                %
\usepackage{import}             %
\usepackage[noend,figure,vlined]{algorithm2e} %
\usepackage{footnote}           %

\usepackage{keycommand}         %

\usepackage{hyphenat}           %

\usepackage{cite}

\usepackage{bibentry}           %

\usepackage{flafter}

\usepackage{dpfloat}

\usepackage{array}              %
\usepackage{longtable}          %
\usepackage{multirow}           %
\usepackage{makecell}           %
\usepackage{tabularx}           %
\usepackage{booktabs}           %

\usepackage{pgffor}             %
\usepackage{adjustbox}          %

\usepackage{circledsteps}       %
\usepackage{nicefrac}

\usepackage{amsmath}            %
\ifIEEE{
    \usepackage{amssymb}            %
    \usepackage{amsfonts}           %
}
\usepackage{amsthm}             %
\usepackage{mathtools}          %

\usepackage{caption}
\usepackage{subcaption}

\usepackage{hyperref}

\usepackage[capitalise,nameinlink,noabbrev]{cleveref}

\usepackage[
    nomain,             %
    acronym,
]{glossaries-extra}

\newignoredglossary{ignored-glossary}

\usepackage{ifdraft}

\usepackage[colorinlistoftodos,prependcaption,textsize=tiny,obeyFinal]{todonotes} %

\newcommand{\fixme}[2][]{\todo[backgroundcolor=magenta,#1]{FIXME: #2}}

\glsenableentrycount
\glssetcategoryattribute{abbreviation}{entrycount}{1}

\setdescription{topsep=0pt,leftmargin=0pt,labelindent=\parindent}

\protected\def\nmicro{\text{\ensuremath{\mu}}}
\glsenableentrycount
\let\gls\cgls
\let\glspl\cglspl
\let\Gls\cGls
\let\Glspl\cGlspl

\glspatchtabularx

\glsdisablehyper

\pgfkeys{/csteps/inner color=white}
\pgfkeys{/csteps/inner xsep=7pt}
\pgfkeys{/csteps/outer color=black}
\pgfkeys{/csteps/fill color=black}
\newcommand\circleref[1]{{\;\smaller\Circled[]{\textbf{#1}}}}

\definecolor{darkgreen}{RGB}{0, 120, 0}

\DeclareSIUnit{\nothing}{\relax}

\glsenableentrycount

\DeclareSIUnit{\pkt}{p}
\DeclareSIUnit{\b}{b}
\DeclareSIUnit{\bits}{bits}
\DeclareSIUnit{\bytew}{byte}
\DeclareSIUnit{\bytes}{bytes}
\DeclareSIUnit{\packets}{packets}

\mathchardef\mhyphen="2D

\newabbreviation[longplural={Network Points of Presence}]{npop}{N-PoP}{Network Point of Presence}
\newabbreviation[longplural={system-on-chips}]{soc}{SoC}{system-on-a-chip}
\newabbreviation[shortplural={CAPEX}]{capex}{CAPEX}{Capital Expenditure}
\newabbreviation[shortplural={IDS}]{ids}{IDS}{Intrusion Detection System}
\newabbreviation[shortplural={OPEX}]{opex}{OPEX}{Operating Expense}

\newabbreviation{3gpp}{3GPP}{3rd Generation Partnership Project}
\newabbreviation{ac}{AC}{access control}
\newabbreviation{adas}{ADAS}{Advanced Driver-Assistance System}
\newabbreviation{adc}{ADC}{analog-to-digital converter}
\newabbreviation{alu}{ALU}{Arithmetic-Logic Unit}
\newabbreviation{amf}{AMF}{Access and Mobility Management Function}
\newabbreviation{api}{API}{application-programming interface}
\newabbreviation{atet}{AT\&T}{American Telephone \& Telegraph}
\newabbreviation{awsecs}{ECS}{Elastic Container Service}
\newabbreviation{bkl}{BKL}{Big Kernel Lock}
\newabbreviation{bpu}{BPU}{Branch Prediction Unit}
\newabbreviation{bt}{BT}{British Telecom}
\newabbreviation{cbs}{CBS}{Constant Bandwidth Server}
\newabbreviation{cfs}{CFS}{Completely Fair Scheduler}
\newabbreviation{cmos}{CMOS}{Complementary Metal–Oxide–Semiconductor}
\newabbreviation{cn}{CN}{Core Network}
\newabbreviation{cots}{COTS}{Commercial Off-the-Shelf}
\newabbreviation{csr}{CSR}{control and status register}
\newabbreviation{cu}{CU}{Central Unit}
\newabbreviation{cu/du}{CU/DU}{Central Unit/Distributed Unit}
\newabbreviation{cupti}{CUPTI}{Cuda Performance Tools Interface}
\newabbreviation{dag}{DAG}{Directed Acyclic Graph}
\newabbreviation{ddio}{DDIO}{Data Direct I/O Technology}
\newabbreviation{dlp}{DLP}{Data-Level Parallelism}
\newabbreviation{dpdk-ans}{DPDK-ANS}{DPDK Accelerated Network Stack}
\newabbreviation{dpdk}{DPDK}{Data Plane Development Kit}
\newabbreviation{dpi}{DPI}{deep packet inspection}
\newabbreviation{dpm}{DPM}{Dynamic Power Management}
\newabbreviation{dpr}{DPR}{Dynamic Partial Reconfiguration}
\newabbreviation{dprio}{DP}{Dynamic Priority}
\newabbreviation{dt}{DT}{Deutsche Telekom}
\newabbreviation{du}{DU}{Distributed Unit}
\newabbreviation{dvfs}{DVFS}{Dynamic Voltage and Frequency Scaling}
\newabbreviation{dvs}{DVS}{vSphere Distributed Switch}
\newabbreviation{eal}{EAL}{Environment Abstraction Layer}
\newabbreviation{edd}{EDD}{Event-Driven Delay-Induced}
\newabbreviation{edf}{EDF}{Earliest Deadline First}
\newabbreviation{eetb}{EETB}{estimated execution time bound}
\newabbreviation{egid}{EGID}{effective group ID}
\newabbreviation{embb}{eMBB}{enhanced Mobile Broadband}
\newabbreviation{enb}{eNB}{Evolved NodeB}
\newabbreviation{epc}{EPC}{Evolved Packet Core}
\newabbreviation{etsi}{ETSI}{European Telecommunications Standards Institute}
\newabbreviation{euid}{EUID}{effective user ID}
\newabbreviation{f1ap}{F1AP}{F1 Application Protocol}
\newabbreviation{f1c}{F1-C}{F1 Control Plane Interface}
\newabbreviation{f1oip}{F1oIP}{F1-over-IP}
\newabbreviation{f1u}{F1-U}{F1 User Plane Interface}
\newabbreviation{fdio}{FD.io}{Fast Data Project}
\newabbreviation{fifo}{FIFO}{First In First Out}
\newabbreviation{first}{FIRST}{Flexible Integrate Real-time Scheduling Technologies}
\newabbreviation{fiveg}{5G}{5th Generation}
\newabbreviation{fpga}{FPGA}{Field Programmable Gate Array}
\newabbreviation{fprio}{FP}{Fixed Priority}
\newabbreviation{fpu}{FPU}{Floating-Point Unit}
\newabbreviation{gedf}{G-EDF}{Global Earliest Deadline First}
\newabbreviation{gnb}{gNB}{Next Generation NodeB}
\newabbreviation{gtp}{GTP}{GPRS Tunnelling Protocol}
\newabbreviation{gui}{GUI}{graphical user interface}
\newabbreviation{hpc}{HPC}{High Performance Computing}
\newabbreviation{hss}{HSS}{Home Subscriber Service}
\newabbreviation{i2c}{I2C}{Inter Integrated Circuit}
\newabbreviation{iaas}{IaaS}{Infrastructure as a Service}
\newabbreviation{ict}{ICT}{Information and Communications Technology}
\newabbreviation{ilp}{ILP}{Integer Linear Programming}
\newabbreviation{iommu}{IOMMU}{Input-Output Memory Management Unit}
\newabbreviation{iot}{IoT}{Internet of Things}
\newabbreviation{ipc}{IPC}{Inter Process Communication}
\newabbreviation{isa}{ISA}{Instruction Set Architecture}
\newabbreviation{isg}{ISG}{Industry Specification Group}
\newabbreviation{itu}{ITU}{International Mobile Telecommunications}
\newabbreviation{itur}{ITU-R}{ITU Radiocommunication Sector}
\newabbreviation{let}{LET}{Logical Execution Time}
\newabbreviation{libos}{LibOS}{Library Operating System}
\newabbreviation{libosnuse}{LibOS-NUSE}{LibOS Network Stack in Userspace} %
\newabbreviation{llc}{LLC}{Last Level Cache}
\newabbreviation{lls}{LLS}{Linear Least Squares}
\newabbreviation{lte}{LTE}{Long Term Evolution}
\newabbreviation{lut}{LUT}{Lookup Table}
\newabbreviation{lxc}{LXC}{Linux Containers}
\newabbreviation{mac}{MAC}{Media Access Control}
\newabbreviation{mano}{MANO}{Management and Orchestration}
\newabbreviation{mde}{MDE}{Model-Driven Engineering}
\newabbreviation{milp}{MILP}{Mixed-Integer Linear Programming}
\newabbreviation{miqcp}{MIQCP}{Mixed-Integer Quadratic Constraint Programming}
\newabbreviation{mme}{MME}{Mobility Management Entity}
\newabbreviation{mmtc}{mMTC}{massive Machine-Type Communications}
\newabbreviation{mp2soc}{MP2SoC}{Massively Parallel Multi-Processors System-on-Chip}
\newabbreviation{napi}{NAPI}{Linux New API}
\newabbreviation{nas}{NAS}{Network Attached Storage}
\newabbreviation{nfv}{NFV}{Network Function Virtualization}
\newabbreviation{nfvi}{NFVI}{Network Function Virtualization Infrastructure}
\newabbreviation{nfvo}{NFVO}{NFV Orchestrator}
\newabbreviation{ng}{NG}{next-generation}
\newabbreviation{ngc}{NGC}{Next-Generation Core}
\newabbreviation{ngran}{NG-RAN}{Next-Generation RAN} %
\newabbreviation{nic}{NIC}{Network Interface Controller}
\newabbreviation{nnls}{NNLS}{Non-Negative Least Squares}
\newabbreviation{nsa}{NSA}{Non-SA} %
\newabbreviation{numa}{NUMA}{Non-Uniform Memory Access}
\newabbreviation{oai}{OAI}{OpenAirInterface}
\newabbreviation{oran}{O-RAN}{Open RAN} %
\newabbreviation{osa}{OSA}{OAI Software Alliance} %
\newabbreviation{ovs}{OVS}{Open vSwitch}
\newabbreviation{paae}{PAAE}{Percentage Average Absolute Error}
\newabbreviation{paas}{PaaS}{Platform as a Service}
\newabbreviation{pcb}{PCB}{printed circuit board}
\newabbreviation{pcc}{PCC}{Pearson Correlation Coefficient}
\newabbreviation{pcp}{PCP}{Priority Ceiling Protocol}
\newabbreviation{pe}{PE}{Processing Element}
\newabbreviation{pedf}{P-EDF}{Partitioned Earliest Deadline First}
\newabbreviation{pf}{PF}{Physical Function}
\newabbreviation{phy}{PHY}{physcal layer}
\newabbreviation{pid}{PID}{process ID}
\newabbreviation{pip}{PIP}{Priority Inheritance Protocol}
\newabbreviation{pl}{PL}{Programmable Logic}
\newabbreviation{pm}{PM}{Poll Mode Driver}
\newabbreviation{pmc}{PMC}{Performance Monitoring Counter}
\newabbreviation{pmd}{PMD}{Poll Mode Driver}
\newabbreviation{pmu}{PMU}{Performance Monitor Unit}
\newabbreviation{pnf}{PNF}{Physical Network Function}
\newabbreviation{ps}{PS}{Processing System}
\newabbreviation{psu}{PSU}{Power Supply Unit}
\newabbreviation{pu}{PU}{processing unit}
\newabbreviation{qos}{QoS}{Quality of Service}
\newabbreviation{ran}{RAN}{Radio Access Network}
\newabbreviation{rapl}{RAPL}{Running Average Power Limit}
\newabbreviation{rdma}{RDMA}{Remote Direct Memory Access}
\newabbreviation{rm}{RM}{Rate Monotonic}
\newabbreviation{rr}{RR}{Round Robin}
\newabbreviation{rrc}{RRC}{Radio Resource Control}
\newabbreviation{rru}{RRU}{Remote Radio Unit}
\newabbreviation{rt}{RT}{real-time}
\newabbreviation{rtkit}{RTKit}{RealtimeKit}
\newabbreviation{rtt}{RTT}{round-trip time}
\newabbreviation{ru}{RU}{Remote Unit}
\newabbreviation{sa}{SA}{Stand Alone}
\newabbreviation{san}{SAN}{Storage Area Network}
\newabbreviation{sba}{SBA}{service-based architecture}
\newabbreviation{sctp}{SCTP}{Stream Control Transmission Protocol}
\newabbreviation{sdn}{SDN}{Software-Defined Networking}
\newabbreviation{simd}{SIMD}{Single Instruction, Multiple Data}
\newabbreviation{sm}{SM}{Streaming Multiprocessor}
\newabbreviation{smf}{SMF}{Session Management Function}
\newabbreviation{spgw}{SPGW}{Serving Gateway/PDN Gateway}
\newabbreviation{sriov}{SR-IOV}{Single-Root I/O Virtualization}
\newabbreviation{srp}{SRP}{Stack Resource Policy}
\newabbreviation{sse}{SSE}{Streaming SIMD Extensions}
\newabbreviation{stream}{STREAM}{Simulation Tool for Energy Efficient Real Time Scheduling and Analysis}
\newabbreviation{svr}{SVR}{Support Vector Regression}
\newabbreviation{tgid}{TGID}{thread group ID}
\newabbreviation{tid}{TID}{thread ID}
\newabbreviation{tif}{TIF}{Top Island First}
\newabbreviation{tlb}{TLB}{Translation Lookaside Buffer}
\newabbreviation{tldk}{TLDK}{FD.io Transport Layer Development Kit}
\newabbreviation{tsc}{TSC}{Time Stamp Counter}
\newabbreviation{ue}{UE}{User Equipment}
\newabbreviation{uio}{UIO}{Userspace I/O}
\newabbreviation{upf}{UPF}{User Plane Function}
\newabbreviation{urllc}{URLLC}{Ultra-Reliable and Low-Latency Communications}
\newabbreviation{vf}{VF}{Virtual Function}
\newabbreviation{vfio}{VFIO}{Virtual Function I/O}
\newabbreviation{vim}{VIM}{Virtual Infrastructure Manager}
\newabbreviation{vm}{VM}{Virtual machine}
\newabbreviation{vmm}{VMM}{Virtual Machine Manager}
\newabbreviation{vnf}{VNF}{Virtual Network Function}
\newabbreviation{vnfc}{VNFC}{Virtual Network Function Component}
\newabbreviation{vnfm}{VNFM}{VNF Manager}
\newabbreviation{vpp}{VPP}{Vector Packet Processing}
\newabbreviation{vran}{vRAN}{Virtualized RAN} %
\newabbreviation{wcet}{WCET}{Worst-Case Execution Time}

\newabbreviation{ann}{ANN}{Artificial Neural Network}
\newabbreviation{gpuperfapi}{GPUPerfAPI}{GPU Performance API}
\newabbreviation{igpu}{iGPU}{Integrated GPU}
\newabbreviation{mape}{MAPE}{Mean Absolute Percentage Error}

\newabbreviation[
  longplural={Operating Systems},
  shortplural={OSes}
]{os}{OS}{Operating System}

\newabbreviation[
  long={\protect\ifglsused{os}{General Purpose OS}{General Purpose \glsdisp[hyper=false]{os}{Operating Systyem}}},
  longplural={\protect\ifglsused{os}{General Purpose OSes}{General Purpose \glsdisp[hyper=false]{os}{Operating Systyems}}},
  short={GPOS},
  shortplural={GPOSes},
]{gpos}{GPOS}{General Purpose Operating System}

\newabbreviation{gpgpu}{GP-GPU}{General Purpose Computing on GPU}

\newabbreviation{fsf}{FSF}{FIRST Scheduling Framework}

\setabbreviationstyle[common]{short-nolong}

\newabbreviation[category=common]{arp}{ARP}{Address Resolution Protocol}
\newabbreviation[category=common]{dhcp}{DHCP}{Dynamic Host Configuration Protocol}
\newabbreviation[category=common]{icmp}{ICMP}{Internet Control Message Protocol}
\newabbreviation[category=common]{ip}{IP}{Internet Protocol}
\newabbreviation[category=common]{nat}{NAT}{Network Address Translation}
\newabbreviation[category=common]{posix}{POSIX}{Portable Operating System Interface for Computing Environments}
\newabbreviation[category=common]{tcp}{TCP}{Transmission Control Protocol}
\newabbreviation[category=common]{udp}{UDP}{User Datagram Protocol}
\newabbreviation[category=common]{gpu}{GPU}{Graphics Processing Unit}

\setabbreviationstyle[retif-component]{long-only-short-only}

\newabbreviation[category=retif]{retif}{ReTiF}{Real-Time Framework}

\newabbreviation[
  category=retif-component,
  type=ignored-glossary,
]{retif-daemon}{ReTiF Daemon}{\emph{ReTiF Daemon}}

\newabbreviation[
  category=retif-component,
  type=ignored-glossary,
]{retif-library}{ReTiF Library}{\emph{ReTiF Library}}

\newabbreviation{ape}{APE}{Absolute Percentage Error}

\newcommand\colornewstuff{n}

\newcommand*{\seeurl}[1]{\footnote{For more info see \url{#1}.}}

\makeatletter
\newcommand{\etc}{etc\@ifnextchar{.}{\xspace}{.\@\xspace}}
\newcommand{\vs}{vs\@ifnextchar{.}{\xspace}{.\@\xspace}}
\newcommand{\etal}{et al\@ifnextchar{.}{\xspace}{.\@\xspace}}
\makeatother

\newcommand\quotes[1]{``{#1}''}

\newcolumntype{L}{X}
\newcolumntype{R}{>{\raggedleft\arraybackslash}X}
\newcolumntype{C}{>{\centering\arraybackslash}X}

\newcolumntype{K}[1]{>{\raggedright\arraybackslash}m{#1}}
\newcolumntype{M}[1]{>{\centering\arraybackslash}m{#1}}
\newcolumntype{H}[1]{>{\raggedleft\arraybackslash}m{#1}}

\newcommand{\ccode}[1]{\mbox{\texttt{#1}}}

\newcommand\ethzstuffkill{n}

\newcommand\ethzstuff[1]{
  \begingroup%
  \if\ethzstuffkill y%
  \else%
    \ifoptionfinal{%
      #1%
    }{%
      \color{green!50!black}%
      #1%
    }%
  \fi%
  \endgroup%
}

\if\ethzstuffkill y
  
\else
  
\fi

\newcommand\mhz{\mega\hertz}

\newcommand\xavierboard{NVIDIA Jetson AGX Xavier\xspace}

\newcommand\fundinginfo{
  This work has received funding from the European Commission through the EU H2020 research project AMPERE (A Model-driven development framework for highly Parallel and EneRgy-Efficient computation supporting multi-criteria optimization) under grant agreement no.\ 871669.
}

\begin{document}

\title{Data-Driven Power Modeling and Monitoring via Hardware Performance Counters Tracking}

\ifACM{\newcommand\myauthor[2][]{
    \pgfkeys{/gara/author#2/.cd,
        name/.initial = {},
        orcid/.initial = {},
        email/.initial = {},
        member/.initial = {},
        affiliation/.initial = {},
        #1
    }
}

\newcommand\myaffiliation[2][]{
    \pgfkeys{/gara/affiliation#2/.cd,
        lab/.initial = {},
        school/.initial = {},
        zip/.initial = {},
        city/.initial = {},
        country/.initial = {},
        #1
    }
}

\newcommand\authorName[1]{\pgfkeysvalueof{/gara/author#1/name}}
\newcommand\authorEmail[1]{\pgfkeysvalueof{/gara/author#1/email}}
\newcommand\authorAffiliation[1]{\pgfkeysvalueof{/gara/author#1/affiliation}}
\newcommand\authorOrcid[1]{\pgfkeysvalueof{/gara/author#1/orcid}}

\newcommand\authorOrcidLink[1]{%
    \ifthenelse{%
        \equal{\authorOrcid{#1}}{}}{}{%
        \,\orcidlink{\authorOrcid{#1}}}}

\newcommand\authorMember[1]{%
    \ifthenelse{%
        \equal{\pgfkeysvalueof{/gara/author#1/member}}{}}{}{%
        ,~\IEEEmembership{\pgfkeysvalueof{/gara/author#1/member}}}}

\newcommand\affiliationLab[1]{\pgfkeysvalueof{/gara/affiliation#1/lab}}
\newcommand\affiliationSchool[1]{\pgfkeysvalueof{/gara/affiliation#1/school}}
\newcommand\affiliationCity[1]{\pgfkeysvalueof{/gara/affiliation#1/city}}
\newcommand\affiliationZip[1]{\pgfkeysvalueof{/gara/affiliation#1/zip}}
\newcommand\affiliationCountry[1]{\pgfkeysvalueof{/gara/affiliation#1/country}}

\myauthor[%
    name={Gabriele~Ara},
    orcid={0000-0001-5663-4713},
    email={gabriele.ara@santannapisa.it},
    affiliation={SantAnna},
]{GAra}

\myauthor[%
    name={Sergio~Mazzola},
    email={smazzola@iis.ee.ethz.ch},
    orcid={0000-0001-8705-8990},
    affiliation={ETHZ},
]{SMazzola}

\myauthor[%
    name={Thomas~Benz},
    email={tbenz@iis.ee.ethz.ch},
    orcid={0000-0002-0326-9676},
    member={Student Member,~IEEE},
    affiliation={ETHZ},
]{TBenz}

\myauthor[%
    name={Bj{\"o}rn~Forsberg},
    email={bjorn.forsberg@ri.se},
    orcid={},
    affiliation={RISE},
]{BForsberg}

\myauthor[%
    name={Luca~Benini},
    email={lbenini@iis.ee.ethz.ch},
    orcid={0000-0001-8068-3806},
    member={Fellow,~IEEE},
    affiliation={ETHZ}, %
]{LBenini}

\myauthor[%
    name={Tommaso~Cucinotta},
    email={tommaso.cucinotta@santannapisa.it},
    orcid={0000-0002-0362-0657},
    member={Member,~IEEE},
    affiliation={SantAnna},
]{TCucinotta}

\myaffiliation[%
    lab={Real-Time Systems Laboratory},
    school={Scuola Superiore Sant'Anna},
    zip={56124},
    city={Pisa},
    country={Italy},
]{SantAnna}

\myaffiliation[%
    lab={Department of Computer Science},
    school={RISE Research Institutes of Sweden},
    zip={164 40},
    city={Kista},
    country={Sweden},
]{RISE}

\myaffiliation[%
    lab={Integrated Systems Laboratory (IIS)},
    school={ETH Zürich},
    zip={8092},
    city={Zürich},
    country={Switzerland},
]{ETHZ}

\myaffiliation[%
    lab={Department of Electrical, Electronic and Information Engineering (DEI)},
    school={University of Bologna},
    zip={40136},
    city={Bologna},
    country={Italy},
]{UniBo}

\newcommand\ACMaffiliation[1]{%
    \affiliation{
        \institution{\affiliationLab{#1}, \affiliationSchool{#1}}
        \city{\affiliationCity{#1}}
        \country{\affiliationCountry{#1}}
        \postcode{\affiliationZip{#1}}
    }
}

\newcommand\ACMauthor[1]{
    \author{\authorName{#1}}
    \ACMaffiliation{\authorAffiliation{#1}}
    \email{\authorEmail{#1}}
    \orcid{\authorOrcid{#1}}
}

\ACMauthor{SMazzola}
\authornote{Both authors contributed equally to this research.}

\ACMauthor{GAra}
\authornotemark[1]

\ACMauthor{TBenz}
\ACMauthor{BForsberg}

\ACMauthor{TCucinotta}

\ACMauthor{LBenini}
\ACMaffiliation{UniBo}

\renewcommand{\shortauthors}{Mazzola, Ara et al.}
}
\ifIEEE{\newcommand\myauthor[2][]{
    \pgfkeys{/gara/author#2/.cd,
        name/.initial = {},
        orcid/.initial = {},
        email/.initial = {},
        member/.initial = {},
        affiliation/.initial = {},
        #1
    }
}

\newcommand\myaffiliation[2][]{
    \pgfkeys{/gara/affiliation#2/.cd,
        lab/.initial = {},
        school/.initial = {},
        zip/.initial = {},
        city/.initial = {},
        country/.initial = {},
        #1
    }
}

\newcommand\authorName[1]{\pgfkeysvalueof{/gara/author#1/name}}
\newcommand\authorEmail[1]{\pgfkeysvalueof{/gara/author#1/email}}
\newcommand\authorAffiliation[1]{\pgfkeysvalueof{/gara/author#1/affiliation}}
\newcommand\authorOrcid[1]{\pgfkeysvalueof{/gara/author#1/orcid}}

\newcommand\authorOrcidLink[1]{%
    \ifthenelse{%
        \equal{\authorOrcid{#1}}{}}{}{%
        \,\orcidlink{\authorOrcid{#1}}}}

\newcommand\authorMember[1]{%
    \ifthenelse{%
        \equal{\pgfkeysvalueof{/gara/author#1/member}}{}}{}{%
        ,~\IEEEmembership{\pgfkeysvalueof{/gara/author#1/member}}}}

\newcommand\affiliationLab[1]{\pgfkeysvalueof{/gara/affiliation#1/lab}}
\newcommand\affiliationSchool[1]{\pgfkeysvalueof{/gara/affiliation#1/school}}
\newcommand\affiliationCity[1]{\pgfkeysvalueof{/gara/affiliation#1/city}}
\newcommand\affiliationZip[1]{\pgfkeysvalueof{/gara/affiliation#1/zip}}
\newcommand\affiliationCountry[1]{\pgfkeysvalueof{/gara/affiliation#1/country}}

\myauthor[%
    name={Gabriele~Ara},
    orcid={0000-0001-5663-4713},
    email={gabriele.ara@santannapisa.it},
    affiliation={SantAnna},
]{GAra}

\myauthor[%
    name={Sergio~Mazzola},
    email={smazzola@iis.ee.ethz.ch},
    orcid={0000-0001-8705-8990},
    affiliation={ETHZ},
]{SMazzola}

\myauthor[%
    name={Thomas~Benz},
    email={tbenz@iis.ee.ethz.ch},
    orcid={0000-0002-0326-9676},
    member={Student Member,~IEEE},
    affiliation={ETHZ},
]{TBenz}

\myauthor[%
    name={Bj{\"o}rn~Forsberg},
    email={bjorn.forsberg@ri.se},
    orcid={},
    affiliation={RISE},
]{BForsberg}

\myauthor[%
    name={Luca~Benini},
    email={lbenini@iis.ee.ethz.ch},
    orcid={0000-0001-8068-3806},
    member={Fellow,~IEEE},
    affiliation={ETHZ}, %
]{LBenini}

\myauthor[%
    name={Tommaso~Cucinotta},
    email={tommaso.cucinotta@santannapisa.it},
    orcid={0000-0002-0362-0657},
    member={Member,~IEEE},
    affiliation={SantAnna},
]{TCucinotta}

\myaffiliation[%
    lab={Real-Time Systems Laboratory},
    school={Scuola Superiore Sant'Anna},
    zip={56124},
    city={Pisa},
    country={Italy},
]{SantAnna}

\myaffiliation[%
    lab={Department of Computer Science},
    school={RISE Research Institutes of Sweden},
    zip={164 40},
    city={Kista},
    country={Sweden},
]{RISE}

\myaffiliation[%
    lab={Integrated Systems Laboratory (IIS)},
    school={ETH Zürich},
    zip={8092},
    city={Zürich},
    country={Switzerland},
]{ETHZ}

\myaffiliation[%
    lab={Department of Electrical, Electronic and Information Engineering (DEI)},
    school={University of Bologna},
    zip={40136},
    city={Bologna},
    country={Italy},
]{UniBo}

\newcommand\IEEEauthor[1]{\authorName{#1}\authorOrcidLink{#1}\authorMember{#1}}
\newcommand\IEEEaffiliation[1]{%
    \affiliationLab{#1}, \affiliationSchool{#1}, \affiliationZip{#1} \affiliationCity{#1}, \affiliationCountry{#1}}

\author{
    \IEEEauthor{SMazzola},
    \IEEEauthor{GAra},
    \IEEEauthor{TBenz},
    \IEEEauthor{BForsberg},
    \IEEEauthor{TCucinotta}, and
    \IEEEauthor{LBenini}

    \thanks{
        \fundinginfo
    }

    \thanks{
        \authorName{SMazzola} and \authorName{GAra} contributed equally to this work.
    }

    \thanks{
        \authorName{SMazzola} and \authorName{TBenz}
        are with the \IEEEaffiliation{ETHZ}
        (e-mail:
        \authorEmail{SMazzola};
        \authorEmail{TBenz}).
    }

    \thanks{\authorName{BForsberg}
        is with the \IEEEaffiliation{RISE}
        (e-mail:
        \authorEmail{BForsberg}).}

    \thanks{
        \authorName{GAra} and \authorName{TCucinotta}
        are with the \IEEEaffiliation{SantAnna}
        (email:
        \authorEmail{GAra};
        \authorEmail{TCucinotta}).
    }

    \thanks{
        \authorName{LBenini}
        is with the \IEEEaffiliation{ETHZ},
        and also with the \IEEEaffiliation{UniBo}
        (e-mail: \authorEmail{LBenini}).
    }
}
}

\newcommand\putabstract{%
  \begin{abstract}%
\glsresetall{}
In the current high-performance and embedded computing era, full-stack energy-centric design is paramount. Use cases require increasingly high performance at an affordable power budget, often under real-time constraints. %
Extreme heterogeneity and parallelism address these issues but greatly complicate online power consumption assessment, which is essential for dynamic hardware and software stack adaptations.
We introduce a novel architecture-agnostic power modeling methodology with state-of-the-art accuracy, low overhead, and high responsiveness. 
Our methodology identifies the best \glspl{pmc} to model the power consumption of each hardware sub-system at each \gls{dvfs} state.
The individual linear models are combined into a complete model that effectively describes the power consumption of the whole system, %
achieving high accuracy and low overhead. Our evaluation reports an average estimation error of \SI{7.5}{\percent} for power consumption and \SI{1.3}{\percent} for energy.
Furthermore, we propose Runmeter, an open-source, \gls{pmc}-based monitoring framework integrated into the Linux kernel. Runmeter manages \gls{pmc} samples collection and manipulation, efficiently evaluating our power models at runtime. With a time overhead of only \SI{0.7}{\percent} in the worst case, Runmeter provides responsive and accurate power measurements directly in the kernel, which can be employed for actuation policies such as \gls{dpm} and power-aware task scheduling.

  \end{abstract}
}

\ifACM{
  \putabstract



\received{2 January 2025} %
\received[revised]{12 March 2009}
\received[accepted]{5 June 2009}

}

\ifIEEE{
  \IEEEtitleabstractindextext{%
    \putabstract
  }
}

\markboth{IEEE Transactions on Computers}%
{Mazzola \MakeLowercase{\etal}: Data-Driven Power Modeling and Monitoring via Hardware Performance Counters Tracking}

\maketitle

\ifIEEE{
  \IEEEdisplaynontitleabstractindextext
  \IEEEpeerreviewmaketitle
}

\glsresetall{}

    \section{Introduction}\label{sec:introduction}

\IEEEPARstart{R}{ecent} years have seen a dramatic evolution in the embedded and real-time computing landscape, with increasingly demanding requirements. Applications striving for ever-higher computing capabilities and energy efficiency are shifting toward heterogeneous computing
platforms~\cite{Quinones20-AMPERE,Cucinotta2023TECS}. \emph{Embedded \gls{hpc}} applications~\cite{Alcaide20} include soft real-time use cases (such as media streaming, virtual and augmented reality)
as well as hard real-time ones (Industry 4.0 and connected factories, modern automotive and aerospace with self-driving, autonomous vehicles).
These areas often demand embedded systems with enhanced sensing, control, and actuation.
To achieve this, embedded devices are typically equipped with hardware accelerators to execute complex, real-time machine-learning models, as well as 5G communication pipelines~\cite{wang2019benchmarking}.

Heterogeneity and massive parallelism are leveraged by hardware designers to integrate increasingly higher computing capabilities on a single die at an affordable energy budget~\cite{hennessy2019new}.
However, since the end of Dennard's scaling, it has been increasingly challenging to match Moore's predictions~\cite{hennessy2019new}, due to several \emph{walls}, from power consumption to memory to hardware overspecialization~\cite{Fuchs2019accelwall}. In current computer architectures, transistor miniaturization causes increased power density and thermal dissipation issues.
With the limits of current silicon technology exposed, pushing for maximum energy efficiency in a dynamic, workload- and operating-condition-aware fashion is paramount to meet the widespread need for high performance within sustainable power budgets~\cite{duranton2021hipeac}.

The de-facto standard to address power efficiency is \gls{dpm}, integrated even in today's simplest embedded systems in the form of \gls{dvfs} and clock gating.
To exploit the full potential of power-manageable designs, the software stack must also be able to perform intelligent adaptation of the power-related knobs based on the available power-related metrics.
This information must be provided at the lowest levels of the software stack, i.e., the OS kernel. As a matter of fact, this
allows the \emph{task scheduler} to perform informed decisions as to the assignment of computing resources to the running processes based on two factors: the current status of the hardware and accurate estimates of the power consumption
and its variations~\cite{mascitti2020dynamic,Mascitti21}.

For such full-stack, energy-aware dynamic adaptations to be effective, however, \textit{accurate}, \textit{fine-grain}, and \textit{responsive} online power measurements are required. The \textit{fine-granularity} property is twofold: a fine-granularity power measurement has decomposability properties, i.e., it is able to provide insights into the power consumption of individual hardware sub-systems~\cite{bertran2012systematic}. This is essential for \gls{dpm}. Additionally, a fine-granularity measurement provides information about task-level power consumption, useful for task scheduling. The \textit{responsiveness} of a power measurement technique refers to its ability to reflect the actual profile of the hardware activity with negligible latency, which requires a low power gauging overhead. This is necessary for the stability of any actuation control loop.
The most commonly adopted solution in this regard consists of analog power sensors. However, as discussed in \Cref{subsec:analog_power}, they do not typically satisfy such requirements~\cite{lin2020taxonomy}.

As an alternative to power sensing, power models have been extensively researched to obtain power measurements better suited for dynamic, online adaptations of hardware and software.
It is well-known that \gls{pmc} activity effectively correlates to power consumption~\cite{bellosa2000benefits}, enabling accurate power modeling for fast, responsive, and fine-grain indirect power gauging~\cite{bertran2012systematic}. However, the complexity of selecting appropriate \glspl{pmc} and understanding the underlying hardware architecture, coupled with the modeling challenges of \gls{dvfs}, complicates their broader application in heterogeneous parallel systems.

This paper introduces a performance-counter-based approach to power consumption estimation for
modern, \gls{dvfs}-enabled, heterogeneous, embedded computing systems, extending our previous work on the topic~\cite{Mazzola2022SAMOS}. We outline a simple yet effective data-driven statistical model for the power consumption of a generalized system by decomposing it into smaller and more easily approachable sub-systems, focusing on individual \gls{dvfs} states. Each sub-system is modeled separately, and, in a later step, the sub-models are re-composed into a single \gls{lut} of models efficiently and accurately describing the system in its entirety. In particular, we target heterogeneous systems composed of CPU hosts and hardware accelerators, like \glsxtrshortpl{gpu} and \glsxtrshortpl{fpga}, that expose \glspl{pmc}. The generality of this approach makes it capable of supporting any additional sub-system that exposes meaningful hardware counters. %

We also propose \emph{Runmeter}, an architecture-agnostic implementation of the model within the Linux kernel that automatizes the collection of \gls{pmc} samples and the online evaluation of the power model in a lightweight and responsive fashion.

The approach is demonstrated and validated on a modern, heterogeneous, \gls{dvfs}-enabled target platform, the NVIDIA Jetson AGX Xavier board. Considering its CPU and GPU sub-systems, our combined, system-level power model achieves an instantaneous power \gls{mape} of \SI{7.5}{\percent} and an energy estimation error of \SI{1.3}{\percent}. On this platform, the run-time implementation in Linux exhibits worst-case overheads of 0.7\%.

With respect to our previously published work~\cite{Mazzola2022SAMOS}, in this paper, we generalize the way sub-systems are treated, abandoning any specificity related to CPU and \gls{gpu} sub-systems, and we present its actual implementation on Linux, evaluating its accuracy,  measurement delay, and run-time overheads.

The rest of this paper is organized as follows.
    \Cref{sec:background} introduces essential background knowledge and justifications for the approach described in this paper.
    \Cref{sec:relwork} frames our contributions into the context of the related work in this field.
    \Cref{sec:model} describes the data-driven approach to the modeling of power consumption of a system,
    as a generalization of \cite{Mazzola2022SAMOS}.
    \Cref{sec:approach} describes the architecture of our novel power monitoring framework integrated within the Linux kernel.
    \Cref{sec:experiments} showcases the evaluation of our power model as a result of both the model construction, offline validation, and online evaluation.
    \Cref{sec:conclusions} contains final remarks and discusses possible future directions in this topic.

\section{Background}
\label{sec:background}

An energy-centric design flow addresses energy efficiency from the point of view of the whole system stack, which is fundamental for complex parallel and heterogeneous systems. To motivate our proposal, we present two main use cases of power-aware dynamic adaptations of the hardware and software, namely \gls{dpm} and power-aware task scheduling, requiring robust online power measurements.
In the following, we also discuss why typical implementations of analog power gauges are not suitable for such hardware and software dynamic adaptations, motivating the choice of \gls{pmc}-based power models.

\subsection{Dynamic Power Management}

\glsreset{dpm}
\emph{\Gls{dpm}} is an essential feature when it comes to increasing performance and energy efficiency through parallelism and heterogeneity~\cite{taylor2012dark}. Having multiple processing elements or computing islands enables fine-grained control of their power status through clock gating and \gls{dvfs}. In this way, hardware portions can be independently turned off or slowed down based on the phase of the running workload. This mechanism
allows modern architectures to keep enough logic on the same die such that hardware overspecialization is prevented while still achieving high energy efficiency.

However, \gls{dpm}-enabled hardware alone is not enough. Actuation policies that are aware of the current power consumption of the hardware must be in place, and they must leverage robust power measurements to close the control loop~\cite{chau2017energy}.

\subsection{Power-Aware Task Scheduling}

\emph{Power-aware task scheduling} integrates \gls{dpm} techniques, such as \gls{dvfs}, with the task scheduler~\cite{Marinoni2016}. Augmenting the process scheduler with intimate knowledge on the tasks that it is managing is paramount for developing intelligent scheduling techniques~\cite{saez2017pmctrack,Rapp2020}. This is especially true for applications characterized by real-time constraints running on embedded systems, where power awareness within the task scheduler is essential~\cite{Balsini2019,mascitti2020dynamic}.

As a first-order approximation, task execution time scales with the CPU frequency, therefore not all the available \gls{dvfs} states might be suitable for meeting the timing constraints of the real-time tasks~\cite{balsini2019modeling}. In addition, by managing the migration of tasks among computational units, the scheduler is an important tool to regulate the thermal behavior of the platform. This is a ubiquitous issue of embedded devices, which can rarely afford an external cooling mechanism~\cite{Ara2022SAC}.

Furthermore, the scheduler decisions are based on currently known system conditions (e.g., its power consumption or expected task finishing time), but their impact might extend for a long time~\cite{Marinoni2016}. A prediction may indeed lead to unrecoverable situations if it is later proved wrong. This problem is exacerbated by the slow reaction time and high access overhead typical of traditional analog power monitors~\cite{Castro2018}, the de facto standard for many embedded systems, critically hindering the capability of the scheduler to make well-informed decisions.

To address these issues and enable future exploitation of effective power-aware scheduling techniques, in \Cref{sec:model,sec:approach}, we propose a methodology to provide the scheduler with accurate power estimates characterized by fast reaction times and lower access overhead compared to analog power monitors.

\subsection{On the Shortcomings of Analog Power Measurements}
\label{subsec:analog_power}

For power-aware dynamic hardware adaptations to be possible, online power measurement is a requirement. To effectively leverage techniques such as \gls{dpm} and power-aware task scheduling, the online power measurement approach must possess the following properties:
\begin{enumerate}
    \item \textit{accuracy}: accurate measurements with high enough resolution and sensitivity are required to feed the actuation control loop properly;
    \item \textit{fine granularity}, both in terms of introspection into hardware sub-system power consumption (i.e., decomposability) and task-level power budgeting;
    \item \textit{responsiveness}: promptly reflect the activity profile of the hardware platform to provide the control loop with minimal-latency power measurements.
\end{enumerate}

Several off-the-shelf \glspl{soc} come equipped with built-in power sensors; their typical implementation consists of analog current and voltage meters, an \gls{adc}, and some additional circuitry that implements communication with the host system.

The reliability of analog power sensors for power-aware dynamic adaptations depends on a large number of factors, which typically do not satisfy the mentioned requirements~\cite{lin2020taxonomy}.
The accuracy of analog power measurements is heavily affected by the platform's \gls{pcb} configuration and further conditioned by the \gls{adc} in terms of resolution.
As an example, for the NVIDIA Jetson AXG Xavier board, the two built-in INA3221 power monitors have a resolution of approximately \SI{200}{\milli\watt} only~\cite{nvidiaxavier,ina3221}.

Due to their nature, analog sensors are not integrated on the same die of the \gls{soc}, hence can rarely provide the level of introspection of individual hardware sub-systems. For the same reason, off-chip parasitics pollute their measurements with longer transients, impacting the responsiveness of their power measures~\cite{Castro2018}. Their latency is further affected by the communication channel with the host, usually implemented by a serial protocol such as \glsxtrshort{i2c}, which does not match the speed of the digital domain.

Moreover, due to their physical size and deployment costs, analog gauges usually showcase inadequate scalability.
Several additional solutions concern the adoption of external, high-accuracy, dedicated measurement devices~\cite{lin2020taxonomy}. However, despite the more precise measurements, such solutions are obviously unsuitable for final deployment due to their cost and even worse integration issues, scalability, and introspection capabilities.

Although unsuitable for direct power measurements, built-in analog sensors do not require external equipment, and they can be programmatically and reliably driven. Therefore, they prove useful to build accurate, fine-grain, and responsive \gls{pmc}-based power models through the approach showcased in \Cref{sec:model}. This is demonstrated in \Cref{sec:results-runmeter}.

\subsection{\glsfmtshort{pmc}-based Power Estimation}

Prior work shows that \glspl{pmc} activity correlates with the power consumption of diverse hardware components~\cite{bellosa2000benefits,bircher2011complete,wang2019statistic}. Being typically accessible via \glspl{csr}, their usage is cheap, and their readings are fast and reliable.
As part of the digital domain, \gls{pmc} activity promptly reflects the current state of the hardware resources, exposing desirable responsiveness properties.
\Glspl{pmc} also provide a high degree of introspection into individual hardware sub-systems~\cite{isci2003runtime}, which empowers \gls{pmc}-based models with a degree of decomposability exactly equivalent to the availability of \glspl{pmc}~\cite{bertran2012systematic}.
Therefore, accurate, fine-grain, and responsive power models, such as the one we propose in \Cref{sec:model}, can be derived from them.

While \gls{pmc}-based power models can achieve the accuracy, fine granularity, and responsiveness required by power-aware dynamic adaptations, a model-based approach poses further challenges to the power estimation.
Modern computer architectures, even those shipped as part of embedded systems, expose hundreds of countable performance events \cite{nvidiaxavier,armcortexA57}. Hence, the parameter selection for a robust statistical power model often requires considerable knowledge of the underlying hardware.
Growing parallelism and heterogeneity, together with the frequent lack of open documentation, further raise the challenge.
A careful choice of the model parameters is necessary for several additional factors:
first, \glspl{pmu} can simultaneously track only a limited number of performance counters~\cite{armcortexA57,pi2019study}. Second, the amount of model predictors directly impacts its evaluation overhead, which must be small for practical \gls{dpm} strategies and minimal interference of the scheduler with regular system operation. \Gls{dvfs} determines an additional layer of modeling complexity, as hardware behavior at varying frequencies has to be considered.

To the best of our knowledge, our approach, proposed in \Cref{sec:model,sec:approach}, is the first one to holistically address all the mentioned challenges.

\section{Related Work}
\label{sec:relwork}
\definecolor{OliveGreen}{rgb}{0.247,0.494,0.192}
\definecolor{YellowOrange}{rgb}{1,0.624,0.169}
\definecolor{OrangeRed}{rgb}{1,0.18,0.345}
\definecolor{Gray}{rgb}{0.6,0.6,0.6}

\colorlet{coloryes}{OliveGreen}
\colorlet{colorjein}{YellowOrange}
\colorlet{colorno}{OrangeRed}

\colorlet{colorverygood}{OliveGreen}
\colorlet{colorgood}{OliveGreen}
\colorlet{colorneutral}{Gray}
\colorlet{colorbad}{OrangeRed}
\colorlet{colorverybad}{OrangeRed}

\newcommand{\verygood}[1]{\textcolor{colorverygood}{#1}}
\newcommand{\good}[1]{\textcolor{colorgood}{#1}}
\newcommand{\neutral}[1]{\textcolor{colorneutral}{#1}}
\newcommand{\bad}[1]{\textcolor{colorbad}{#1}}
\newcommand{\verybad}[1]{\textcolor{colorverybad}{#1}}

\newcommand{\tworows}[2]{\begin{tabular}[c]{@{}c@{}}#1\\ #2\end{tabular}}
\newcommand{\tworowsl}[2]{\begin{tabular}[c]{@{}l@{}}#1\\ #2\end{tabular}}
\newcommand{\relwork}[3]{\vspace{1mm}\tworowsl{#1 \etal}{(#2) #3}}

\newcommand\circledsym[2]{%
  \adjustbox{height=1.15em,margin*=0 -1.25 -2.5 0}{%
    \tikz\node[circle,color=white,fill=#1,inner sep=.2pt,font=\bfseries]{#2};%
  }
}

\newcommand{\yes}{\circledsym{coloryes}{$\pmb\checkmark$}}
\newcommand{\no}{\circledsym{colorno}{$\pmb\times$}}
\newcommand{\jein}{\circledsym{colorjein}{$\pmb\approx$}}
\newcommand{\unknown}{\circledsym{colorneutral}{?}}

\newcolumntype{Y}{>{\centering\arraybackslash}X}
\newcolumntype{Z}{>{\raggedleft\arraybackslash}X}
\newcolumntype{R}{%
  >{\adjustbox{right=7.4em,angle=310,lap=-\width+1.5em}\bgroup}%
  c%
  <{\egroup}%
}
\newcolumntype{M}{%
  >{\adjustbox{right=7.4em,angle=310,lap=-\width+2.5em}\bgroup}%
  c%
  <{\egroup}%
}
\newcommand*\rot{\multicolumn{1}{R}}%
\newcommand*\rotm{\multicolumn{1}{M}}%
\newcommand{\tblrottitle}[1]{\rot{{#1}}}
\newcommand{\tblrottitlem}[1]{\rotm{{\makecell[cr]{#1}}}}

\newcommand{\underlinecenter}[2]{%
\setul{3pt}{.4pt}%
\ul{\mbox{\hspace{#1}}#2\mbox{\hspace{#1}}}}

\begin{table}[t]
  \caption{Comparison between representative works in the literature of \glsfmtshort{pmc}-based power modeling.}
  \label{tab:soa}
  \vspace{-0.4cm} %
  \setlength{\tabcolsep}{0pt}%
  \sisetup{range-phrase=--}%
  \center%
  \begin{tabularx}{\columnwidth}{@{}lYYYYYYYYY@{}}
    \toprule
        \parbox[t][5em][b]{1em}{}
      & \tblrottitle{Heterogeneity}
      & \tblrottitlem{Generality}
      & \tblrottitlem{Automation}
      & \tblrottitlem{Architecture-\\agnostic}
      & \tblrottitlem{Lightweight\\model}
      & \tblrottitlem{DVFS support}
      & \tblrottitlem{Decomposability}
      & \tblrottitlem{Runtime\\monitoring}
      & \tblrottitle{Accuracy (MAPE)}
    \\ %
    \midrule
      \relwork{Bertran}{2012}{\cite{bertran2012systematic}}
      & \tworows{CPU}{only} %
      & \jein{} %
      & \jein{} %
      & \no{} %
      & \no{} %
      & \yes{} %
      & \yes{} %
      & \no{} %
      & \tworows{power}{6-20\%} %
    \\ %
      \relwork{Walker}{2016}{\cite{walker2016accurate}}
      & \tworows{CPU}{only} %
      & \tworows{ARM}{cores} %
      & \yes{} %
      & \no{} %
      & \yes{} %
      & \yes{} %
      & \no{} %
      & \no{} %
      &  \tworows{power}{3-4\%} %
    \\ %
      \relwork{Wang}{2019}{\cite{wang2019statistic}}
      & \tworows{iGPU}{only} %
      & \no{} %
      & \no{} %
      & \no{} %
      & \yes{} %
      & \no{} %
      & \no{} %
      & \no{} %
      & \tworows{power}{3\%} %
    \\ %
      \vspace{1.5mm}\tworowsl{Mammeri et}{al. (2019) \cite{mammeri2019performance}}
      & \yes{} %
      & mobile %
      & \jein{} %
      & \jein{} %
      & \no{} %
      & \no{} %
      & \no{} %
      & \no{} %
      & \tworows{power}{4.5\%} %
    \\ %
      \vspace{1.5mm}\tworowsl{Tarafdar et}{al. (2023) \cite{tarafdar2023power}}
      & \no{} %
      & \tworows{data}{centers} %
      & \unknown{} %
      & \no{} %
      & \yes{} %
      & \unknown{} %
      & \no{} %
      & \yes{} %
      & \tworows{power}{4.7\%} %
    \\ %
        \textbf{This work}
      & \yes{} %
      & \yes{} %
      & \yes{} %
      & \yes{} %
      & \yes{} %
      & \yes{} %
      & \yes{} %
      & \yes{} %
      & \begin{tabular}[c]{@{}c@{}}{\scriptsize power}\\ {\scriptsize 7.5\%}\\ {\scriptsize energy}\\ {\scriptsize 1.3\%}\end{tabular} %
    \\ %
    \bottomrule
  \end{tabularx}
\end{table}

\cite{ahmad2017survey,lin2020taxonomy,Zoni2023Survey} present a comprehensive taxonomy and comparison of different modeling approaches and runtime power monitors, ranging from embedded and edge devices to data centers.

In the following, we mainly focus on the research works related to \gls{pmc}-based power modeling and online, model-based power monitoring, comparing them with our approach. We summarize the key characteristics in \Cref{tab:soa}.

\subsection{\glsfmtshort{pmc}-Based Power Modeling}

\Gls{pmc}-based statistical power models have been a hot research topic for the last 20 years, spanning all computing domains from embedded computing devices at the edge to data centers in the cloud.

\glsreset{rapl}

Bertran~\etal~\cite{bertran2013counter} identify two families of \gls{pmc}-based power models, based on their construction: \emph{bottom-up} and \emph{top-down}.
\emph{Bottom-up} approaches rely on extensive knowledge of the underlying architecture to estimate the power consumption of individual hardware sub-systems.
While the pioneering works of this field fall in this category~\cite{isci2003runtime,bertran2012systematic}, their results highlight the strict dependency between the accuracy of bottom-up models and architectural knowledge of the platform, which jeopardizes the generality of this approach. Even more recent works in this category lack in general applicability. A notable example of these works is Phung \etal~\cite{Phung2020}, which derive a very accurate power model based on a combination of event counters, CPU operating frequencies and even CPU temperature. The general applicability of this approach is hindered by the fact that it requires extensive knowledge of the selected Intel architecture, achieving the best results only leveraging Intel \gls{rapl}~\cite{David2010}.

\sbox0{\gls{rapl}}

\emph{Top-down} approaches target simple, low-overhead, and more generally applicable models. They commonly assume a linear correlation between generic activity metrics and power consumption. %
Among the first attempts in the field, Bellosa~\cite{bellosa2000benefits} models the power consumption of a Pentium II processor with a few manually selected \glspl{pmc}. Subsequent works~\cite{pusukuri2009methodology,singh2009real} refine the idea with a more elaborate procedure for \gls{pmc} selection and multi-core support. Bircher and John~\cite{bircher2011complete} are the first to go towards a thorough system-level power model, tackling the issue from a top-down perspective for each sub-system.
No past research investigates a combination of accurate and lightweight models addressing \gls{dvfs} without requiring expert architectural knowledge, which is instead the subject of this paper.

More recent works target CPU power modeling in mobile and embedded platforms. Walker~\etal~\cite{walker2016accurate} employ a systematic and statistically sound technique for \gls{pmc} selection and train power models for the ARM A7 and A15 embedded processors. However, only one trained weight is used to predict the power consumption at any \gls{dvfs} state, which can lead to large inaccuracies. In contrast to their work, we target a broad range of platforms by composing simpler statistical models, nevertheless carefully dealing with \gls{dvfs} and reaching comparable accuracy numbers.

Top-down power modeling approaches are also applied to \glspl{gpu}.
Wang~\etal~\cite{wang2019statistic} analyze the power consumption of an AMD \gls{igpu}, carefully studying its architecture and selecting the best \glspl{pmc} to build a linear power model. Redundant counters are discarded to reduce the model overhead, with power \gls{mape} below \SI{3}{\%}. However, the generated model is not generally applicable and requires expert knowledge.
Recent works also resort to deep learning for creating accurate black-box power models: Mammeri~\etal~\cite{mammeri2019performance} train an \gls{ann} with several manually chosen CPU and \gls{gpu} \glspl{pmc}. With an average power estimation error of \SI{4.5}{\%}, they report power estimates 3$\times$ more accurate than a corresponding linear model; however, the overhead of evaluating a neural network at runtime is not negligible, as it requires a number of multiply-accumulate operations two orders of magnitude higher than for a linear model. Potentially long training time, complex manual selection of the neural network topology, risk of overfitting, and lack of decomposability are additional drawbacks of this approach.

Tarafdar~\etal~\cite{tarafdar2023power} also propose several power modeling techniques based on multi-variable linear regression, \gls{svr}, and \gls{ann}. While their approach is statistically sound and, in theory, applicable to any generic computing platform, they conceive it as a solution for data centers. Therefore, no fine-grain power information about the computing platform is made available. Moreover, the model parameter selection happens a priori and is not correlated to the modeled platform. Their best power model, which features an evaluation overhead in the order of the microseconds, shows an average power estimation error of at most 4.7\%.

Our proposed model shares its decomposability and responsiveness with bottom-up approaches but resorts to top-down modeling for individual sub-systems: we trade a lower per-component introspection for a systematic modeling procedure requiring very little architectural knowledge and minimal human intervention, targeting broad applicability. We indeed refine the approaches in~\cite{pusukuri2009methodology,bircher2011complete,pi2019study} to allow the selection of a minimal set of best \glspl{pmc} for accurate and lightweight power models.
In addition, we address the platform heterogeneity and \gls{dvfs} capabilities by introducing a \glsxtrshort{lut}-based approach that employs individual, lightweight linear models for each sub-system and for each \gls{dvfs} state. This is particularly suitable for implementation as part of an \gls{os} kernel, thanks to its low overhead when evaluated at runtime.

\subsection{Online Model-based Power Monitoring}

Various tools have been proposed for online, model-based estimation of power consumption through \gls{pmc} sampling.
Most of them implement runtime monitoring by simply sampling the \glspl{pmc} with a fixed periodicity~\cite{Pricopi2013,Rodrigues2013,walker2016accurate}. Some authors highlight that a careful choice of the sampling period is critical to achieving appropriate trade-offs between accuracy and responsiveness~\cite{Rodrigues2013}. Other works deal with the constraint that only a limited number of counters can be activated at any given time, thus resorting to counter multiplexing~\cite{Joseph2001}.
Additionally, some works attempt to use the \glspl{pmc} to attach fine-grain power consumption estimates to CPU sub-components~\cite{isci2003runtime}.

One of the most popular open-source tools for online \gls{pmc} sampling is PMCTrack, developed by Saez~\etal~\cite{saez2017pmctrack}. It replaces the \ccode|perf| utility provided by mainline Linux, which can be accessed both from user space and from the Linux kernel,
in particular from the task scheduler. PMCTrack can monitor per-system, per-CPU, and per-process \glspl{pmc}.
However, it focuses on providing access to \gls{pmc}-based statistics for general-purpose use cases. Therefore, some aspects of its implementation are not suited for
{\em real-time} tasks, that may possess different requirements from those of \ccode{SCHED\_OTHER} tasks. As an example, PMCTrack may delay the generation of a \gls{pmc} sample related to a task until it is chosen again by the task scheduler for being executed. This delay could be detrimental to the responsiveness of a power-monitoring tool.

The implementation of our modeling technique
in the Linux kernel, presented in \Cref{sec:approach}, is a fork of the PMCTrack codebase, re-engineered with the goal of providing a responsive and reliable mechanism to monitor the evolution of \glspl{pmc} and, consequently, the power consumption of the platform, in use-cases that include real-time tasks.
Our implementation employs a mechanism based on a moving sampling window, that does not suffer from the above-mentioned low responsiveness of PMCTrack as estimates are updated at a configurable periodicity. This results in responsive power readings that do not sacrifice the accuracy of the estimates, which depends on the window size~\cite{Rodrigues2013}.

Our modeling approach includes an offline optimal selection of the counters to be employed, which are then set up in the online monitor to get accurate power estimations without requiring counter multiplexing~\cite{Joseph2001}. Through the described sampling of \glspl{pmc}, our online monitoring framework allows the collection of power estimates with sub-system-level introspection, such as individual CPU islands, and at the granularity of individual tasks.

\section{Data-Driven Power Modeling}
\label{sec:model}

This section describes our proposed automated, data-driven pipeline for the training of a \gls{dvfs}-aware power estimation model for heterogeneous platforms. While mostly based on the approach presented in~\cite{Mazzola2022SAMOS}, this work introduces a further generalization that expands its applicability from systems composed of only a CPU and a GPU subsystem to any system decomposable into multiple sub-systems.

Given a generic target platform composed of one or multiple
sub-systems, we provide a systematic and architecture-agnostic approach to its characterization (i.e., to model parameter selection) based on extensive profiling of the exposed \glspl{pmc}. This results in an accurate, responsive, and low-overhead power model for the entire platform and its individual sub-systems.

\begin{figure*}[tp]
    \centering

\newlength{\figurewidth}
\ifACM{\setlength{\figurewidth}{\linewidth}}
\ifIEEE{\setlength{\figurewidth}{.8\linewidth}}

    \includegraphics{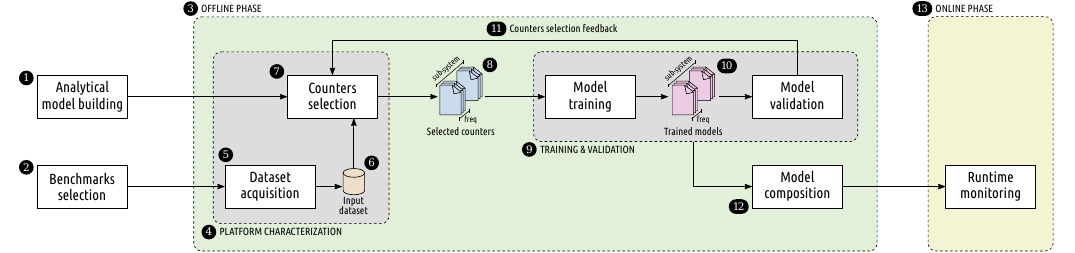}
    \caption{Scheme of the proposed data-driven, automatic power modeling approach for \glsfmtshort{dvfs}-enabled heterogeneous platforms.}
    \label{fig:pipeline}
\end{figure*}

\newcommand\circlefigref[1]{\Cref{fig:pipeline},\circleref{#1}\xspace}
\newcommand\circlefigreftwo[2]{\Cref{fig:pipeline}, \circleref{#1} and \circleref{#2}\xspace}

\subsection{The systematic, data-driven methodology}

For the purpose of this section, we consider a generic computing platform composed of a set $D$ of individual sub-systems $d$. Each sub-system $d$ has a set $F_d$ of possible \gls{dvfs} states, each one characterized by an operating frequency $f_d$.
We define $D^\ast \subseteq D$ as the subset of sub-systems that we target for power modeling. Furthermore, for each $d \in D^\ast$, we define $F_d^\ast \subseteq F_d$ as the subset of $d$'s \gls{dvfs} states that we consider. Both $D^\ast$ and $F_d^\ast$ are user-defined parameters that might vary based on the use-case.

In addition, within a single sub-system $d$, there might be multiple units $i$ running at the same frequency $f_d$, that we would like to track and model individually (e.g., different CPU cores of the same CPU). For the purpose of this discussion, we assume that, for each sub-system's units, up to $N_d$ distinct \glspl{pmc} can be tracked at the same time. For example, typical ARM CPUs simultaneously expose up to four or six individual counters, some of which may be freely selectable by the \gls{os}, while others are fixed.
In the following, we refer to an individual counter exposed by the unit $i$ of the sub-system $d$ with $x_{ij}$, with $j = 1\,..\,N_d$. In general, the set of all $x_{ij}$ of a sub-system $d$ operating at $f_d$ represents the set of input independent variables, or the \emph{predictors}, of our models.

\glsreset{lut}

Given our focus on heterogeneity and \gls{dvfs} capabilities, we deem it impractical to obtain a single mathematical model that would accurately describe the relationship between the power consumption of the platform and \gls{pmc} samples in each possible \gls{dvfs} state, for any hardware sub-system.
For this reason, our power modeling approach relies on the construction of a \gls{lut} able to describe the power consumption of the entire system with high accuracy and low overhead, effectively grasping the different platform behaviors at varying operating frequencies. The \gls{lut} comprises several simpler linear models, one per sub-system $d$ and \gls{dvfs} state $f_d$ that the sub-system can select. Each entry of the \gls{lut} is a linear power model $P_{d}(X_{d, f_{d}}, W_{d, f_{d}})$. The model is driven by the set of \glspl{pmc} $X_{d, f_d}$ tracked when operating at $f_d$, weighted by the set of trained weights $W_{d, f_{d}}$:
\begin{equation}\label{eq:lut}
    LUT[d,f_d] =
    P_{d}(X_{d, f_{d}}, W_{d, f_{d}})
    \quad \textrm{for} \  d \in D^\ast, \; f_d \in F_d^\ast
\end{equation}

\noindent
The goal of our approach, depicted in \Cref{fig:pipeline}, is to obtain all the parameters and weights necessary to construct the set of individual models of the $LUT$.

This starts from the definition of a mathematical expression for a \gls{pmc}-based linear model for each sub-system $d$. Extensive profiling of the platform is then carried out to define the set of model parameters for each sub-system at each frequency. The models obtained in such a way are trained and validated individually and subsequently composed into the final \gls{lut}. Note that the whole procedure can be easily automated and architecture-agnostic, as long as access to representative \glspl{pmc} for each sub-system is granted.

\subsection{Analytical Model Building \& Benchmarks Selection}
\label{subsec:model-benchmarks}

Before the offline modeling phase (\circlefigref{3}), two initial steps are required.
Each sub-system $d$ is associated with a linear power model, which will be trained using linear regression. In the \emph{analytical model building} step (\circlefigref{1}), we define such an expression $P_{d}$ for each sub-system $d$. Thanks to the \gls{lut}-based approach, the frequency $f_d$ is factored out. Therefore, the individual power models are reduced to linear combinations of the \gls{pmc} samples and the trained weights.
\begin{gather}\label{eq:model}
    P_{d}(X_{d, f_{d}}, W_{d, f_{d}}) = L_d + \sum_{i=1}^{\#units} \sum_{j=1}^{N_i} \Big(\frac{1}{T}\cdot x_{ij}\Big)\cdot w_{ij} \\
    \textrm{for} \  d \in D^\ast, \; f_d \in F_d^\ast \ \textrm{and} \ \; L_d, w_{ij} \in W_{d, f_{d}}, \; x_{ij} \in X_{d, f_{d}} \nonumber
\end{gather}
The weight $L_d$ is used to capture the constant component of the power consumption, i.e., the leakage, while the \gls{pmc}-dependent terms vary based on the hardware
activity, modeling the dynamic power.
The factor $\nicefrac{1}{T}$ normalizes the raw \gls{pmc} samples $x_{ij}$ with respect to the sampling period $T$ of the dataset traces: training the model on \gls{pmc} rates rather than absolute values solves sampling jitter and simplifies time-rescaling.

All subsequent steps of the procedure are also based on the careful choice of a representative set of workloads for the heterogeneous platform (\circlefigref{2}). To obtain a proper input dataset, complete coverage of all targeted sub-systems is required to address the heterogeneity of the platform fully. Knowledge of the specific target architecture is nonetheless not required, as high-level considerations about any generic computer architecture suffice for this step.

Secondly, for each sub-system, the workloads should be diverse enough to induce a broad range of different behaviors for a dataset-independent result. The selected benchmarks are employed to build a dataset (\circlefigref{6}) for platform characterization, model training, and model validation. Such a dataset contains the activity traces of the workloads (i.e., the \gls{pmc} traces), each with its associated instantaneous power consumption. The decomposability of the final power model can extend only as in-depth as the available \glspl{pmc} and power measurements.

\subsection{Platform Characterization}
\label{sec:model-characterization-steps}

Given the $P_{d}$ mathematical model for each sub-system $d$, we define \emph{platform characterization} the process of model parameter selection (\circlefigref{4}). In other words, with the platform characterization, we define the set $X_{d,f_d}$ of \glspl{pmc}, which will be used to model the dynamic power of each sub-system $d$ at each frequency $f_d$.

Individually for each sub-system $d$ and frequency $f_d$, we perform a one-time correlation analysis between all of its local \glspl{pmc} and the sub-system power consumption, looking for the optimal $X_{d, f_{d}}$ set in terms of model accuracy, overhead and compliance with \gls{pmu} limitations.
This characterization process is meant as an automatic, architecture-agnostic, and data-driven alternative to manual \gls{pmc} selection, which typically requires architecture-specific considerations.
Given the sub-system $d$, its characterization involves the following steps:

{
\newcommand\mfootnotetext{
Given a set of performance events that cannot be thoroughly profiled simultaneously, we replay the same workload (i.e., we induce the same power consumption profile) until all events have been tracked. This is in contrast to \emph{counter multiplexing}, where the \glspl{pmc} are round-robin time-multiplexed. Counter multiplexing
increases overhead and decreases accuracy, therefore, it is not optimal for online usage.}

\begin{enumerate}
    \item for each \gls{dvfs} state $f_d \in F_d^\ast$, profile \emph{all} performance events exposed by $d$ while tracing $d$'s power; to track all events despite the \gls{pmu} limitations, we make use of \emph{multiple passes}\footnote{\mfootnotetext}~\cite{may2001mpx} (\circlefigref{5});

    \item normalize the \gls{pmc} samples with respect to the sampling periods;

    \item compute a \gls{lls} regression of each \gls{pmc}'s activity trace over its related power measurements, for each $f_d$; events with a p-value above \textit{0.05} are discarded as not reliable for a linear correlation;

    \item individually for each $f_d$, sort the remaining events by their \gls{pcc} and select the best ones that can be profiled simultaneously, i.e., compose $X_{d, f_{d}}$ for all $f_d \in F_d^\ast$ (\circlefigreftwo{7}{8}).
\end{enumerate}
}

While multiple passes are used to profile all available events during the offline platform characterization, with an online deployment in mind for our models, we require $X_{d, f_{d}}$ only to contain events that can be tracked simultaneously by the \gls{pmu}.
To this end, it is usually enough to select, for each frequency $f_d$, the desired number of best counters up to the \gls{pmu} limit.
Finally, the optimal number of performance counters with respect to model accuracy can be found by iteratively employing the estimation error results from the model evaluation step (\circlefigref{11}).

On the other hand, some platforms~\cite{may2001mpx} might have stricter \gls{pmu} constraints. Not only are \glspl{pmc} limited in number, but some of them might be mutually exclusive: \emph{compatibility} must also be considered. As documentation about performance events compatibility is not usually available, we tackle this issue with a \gls{pmu}-aware iterative algorithm based on the profiling \gls{api} provided by the vendor.
Given the sub-system $d$, its frequency $f_d$ and the list of \glspl{pmc}, we heuristically group the \glspl{pmc} with highest \gls{pcc} one by one, adding to $X_{d, f_{d}}$ only events that, based on the provided \gls{api}, can be counted in a single pass.

\subsection{Training, Validation, and System-level Model}
\label{sec:model-train-validation-composition}

With the sets of counters $X_{d, f_{d}}$ from platform characterization, we compose the \gls{lut} of \Cref{eq:lut} by individually training the linear power model $P_{d}(X_{d, f_{d}}, W_{d, f_{d}})$ of each sub-system $d \in D^\ast$ for each $f_d \in F_d^\ast$ (\circlefigref{9}). The output of each training is a set of weights $W_{d, f_d}$ \circleref{10}.
To train each individual $P_{d}(X_{d, f_{d}}, W_{d, f_{d}})$, we perform a \gls{nnls} linear regression of the \glspl{pmc} rates
over the power measurements, obtaining the set of non-negative weights $W_{d, f_{d}}$. Compared to unconstrained \gls{lls}, non-negative weights are physically meaningful and prove to be robust to multicollinearity, which makes our simple models less prone to overfitting.
We subsequently validate each individual $P_{d}(X_{d, f_{d}}, W_{d, f_{d}})$.

After individual training and validation, we combine all the individual sub-system models (\circlefigref{10}) into a system-level power model (\circlefigref{12}) defined as:
\begin{equation}
    \label{eq:sysmodel}
    P_{SYS} = \sum_{d\in D^\ast} P_{d}(X_{d, f_{d}}, W_{d, f_{d}})
\end{equation}
In other words, we fix a $f_d$ for the model of each sub-system $d$. The system-level power model is then the \emph{reduction sum} of the \gls{lut} from \Cref{eq:lut} along the sub-systems dimension $d$.
Modeling the sub-systems individually until this step relieves us from profiling all possible combinations of sub-systems' frequencies, which is simpler, faster, and more robust to overfitting. The final model is decomposable, accurate, and, due to its linearity, computationally lightweight.

The complete model can finally be used online (\circlefigref{13}) to monitor the instantaneous power consumption of the entire system. In order to do so, it is enough to keep available at runtime the weights $W_{d, f_{d}}$ for each $(d,\,f_d)$ combination of interest, acquire the \gls{pmc} measures for the set of model parameters $X_{d, f_{d}}$, and compute the few required multiply-accumulate operations to evaluate the model. A proposed framework for this operation is the object of the next section.

\newcommand\runmeter{Runmeter\xspace}
\newcommand\runmeterframework{\runmeter Framework\xspace}
\newcommand\runmetermodule{\runmeter Kernel Module\xspace}
\newcommand\runmeterpatch{\runmeter Kernel Patch\xspace}

\section{Online Monitoring and Kernel Support}
\label{sec:approach}

The modeling approach discussed in \Cref{sec:model} results in a fast and thorough power model that can provide accurate and introspective power estimates with low overhead. These features are ideal for applications requiring power awareness and real-time capabilities, such as resource allocation and task scheduling.
To demonstrate the applicability of the proposed power modeling methodology, we implement an integrated power monitoring framework for the Linux kernel, which we name \runmeter.
As a case study for such a framework, we implement power estimation and monitoring of the CPU sub-system. However, the implementation can be extended to support additional sub-systems, such as the GPU.

\runmeter is a framework designed to support the runtime estimation of platform- and task-level metrics, including power and energy consumption, through \gls{pmc} tracking. As such, it provides the actual means to feed the \gls{pmc} values to the model presented in Section \ref{sec:model}, forwarding its energy estimation to the Linux scheduler.
\runmeter's modular design makes it highly architecture-agnostic. Only a minimal subset of its components must be re-implemented to support different target platforms. In addition, its flexibility allows easy extension for further relevant metrics to be collected at runtime, transparently to the target architecture.

The \runmeterframework consists of two main components:
\begin{enumerate*}
    \item \runmeterpatch, a patch for the Linux kernel that enables the dynamic loading of the framework into the kernel at runtime;
    \item \runmetermodule, a dynamically loadable Linux kernel module that implements the bulk of \runmeter functionality.
\end{enumerate*}
Additionally, a set of user-space tools is provided to simplify the interaction with the kernel module for configuration and periodical monitoring purposes. In the following, we focus on the description of the in-kernel components.

\subsection{\runmeter Kernel Components}
\label{subsec:runtime-patch}
\label{subsec:runtime-module}

The \runmeterpatch applies minimal changes to the Linux kernel internals, which are otherwise inaccessible from loadable modules.
The patch injects callbacks to \runmeter in key moments of task execution (namely during the context switch and the scheduler tick callbacks), which allows for collecting \glspl{pmc} information as required.

Once loaded into the kernel, the \runmetermodule attaches to the callbacks to trace and collect running statistics on a selection of the available \glspl{pmc}. It does so by taking over \glspl{pmc} management from the Linux kernel and configuring the platform to track the counters according to the result of the platform characterization (\Cref{sec:model-characterization-steps}). Since a different set of counters $X_{\text{CPU},f_{\text{CPU}}}$ can be selected to model the evolution of the platform depending on the \gls{dvfs} state $f_{\text{CPU}}$, the module selects the correct \glspl{pmc} to track accordingly to the model \gls{lut}.
The module also subscribes to the CPU frequency governor (\ccode{CPUFreq}) to be notified of each change of frequency so that it can dynamically reconfigure the set of tracked \glspl{pmc} for each CPU core.

When tracking is enabled, the kernel module generates a new \gls{pmc} sample on each CPU core whenever one of the following events occurs:
\begin{itemize}[noitemsep,nolistsep]
    \item a context switch, in which case a new sample is always generated, or
    \item a user-configurable number of scheduler ticks since the last sample was produced on that core.
\end{itemize}
The first trigger is necessary so that the \gls{pmc} statistics of each individual task can be queried. This allows \runmeter to collect \gls{pmc} information and derive metrics with task-level granularity.
The second trigger, on the other hand, provides an upper bound to the inter-arrival time between two consecutive \gls{pmc} samples. This guarantees that tasks hogging the CPU do not interfere with the monitoring. Since this bound is expressed in terms of scheduler ticks, its granularity depends on the \ccode{CONFIG\_HZ} Linux kernel option.

Proper selection of this upper bound is key to ensuring the desired responsiveness when monitoring CPU counters evolving over time. The basic \gls{pmc} sampling mechanism provides the accumulated value of the event counter since its last reading.
Therefore, a small sampling window might negatively impact the information collected by the \glspl{pmc}: since each sample is tied to a single task, it is difficult to derive any meaningful data about the overall platform status by considering exclusively a single \gls{pmc} sample.
On the other hand, if such an interval is long, the read-out value is not updated often. This is
detrimental for actuation policies requiring high responsiveness~\cite{Rodrigues2013}.

As a trade-off, we devise a \emph{moving-window} approach that decouples the \gls{pmc} sampling period from their observation window. The moving window allows us to obtain \gls{pmc} statistics accumulated over an arbitrarily long window and updated at an arbitrarily fine time granularity.
To implement the moving-window approach, we instantiate a \emph{window buffer} for each \gls{pmc}. Each buffer stores a user-configurable number of \gls{pmc} samples. Whenever a new sample is produced, the window shifts forward.
The value of each \gls{pmc} over the whole window is tracked by summing up all samples in the buffer. This information is updated each time the window moves forward (i.e., a new sample is available). We refer to this value as \emph{synthetic \gls{pmc} sample}.

Consuming synthetic samples provides more meaningful \gls{pmc} data for the metrics to be estimated. However, it comes at the cost of the additional processing for the update of each \gls{pmc}'s moving window. Nevertheless, in \Cref{sec:experiments}, we report a negligible overhead.

\subsection{In-Kernel CPU Power Model}
\label{subsec:cpu_model}

Given the dense stream of synthetic \gls{pmc} samples generated by the core of the \runmeterframework, components like an online CPU power (or energy) monitor are required to re-evaluate their estimates at each update of the corresponding synthetic samples. A high degree of responsiveness in the monitoring, useful for prompt actuation, is achieved when the model evaluation time can keep up with the stream of synthetic samples. In \Cref{sec:model}, we present a power modeling approach devised with online usage in mind, which retains high accuracy despite its low computational complexity.

As a case study, the \runmeterframework implements support for online monitoring of the instantaneous power consumption of the CPU sub-system.
The CPU power monitor in \runmeter implements, for each CPU \gls{dvfs} state, the following model, based on \Cref{eq:model}:
\begin{equation}\label{eq:cpu_power_model}
\begin{aligned}
    P_{\text{CPU}} = L_{\text{CPU}} + \sum_{i = 1}^{\#cores} \sum_{j = 1}^{N_\text{CPU}} \Big(\frac{1}{T'} \cdot x_{ij}\Big)\cdot w_{ij}
\end{aligned}
\end{equation}
\noindent
The weights $L_{\text{CPU}}$ and $w_{ij}$ are fractional values. Therefore, the estimation of the instantaneous power $P_{\text{CPU}}$ at each query of the monitoring framework encompasses fractional additions and multiplications. However, using floating-point arithmetic within the Linux kernel is problematic and expensive.
For this reason, we use fixed-point arithmetic to implement the in-kernel power model.
This decision is supported by the negligible loss of dynamic range and precision with respect to floating-point data type, which we evaluate in \Cref{sec:results-runmeter}.

The factor $\nicefrac{1}{T'}$ normalizes the value of each synthetic sample with respect to the width of the user-configured observation window $T'$. $T'$ might indeed differ from the sampling period $T$ of the model training dataset (\Cref{eq:model}).
Thanks to the linearity of our models, we can perform the normalization by multiplying only once the result of the summations. This achieves arbitrary time-rescaling with negligible overhead.

\begin{equation*}
\begin{aligned}
    P_{\text{CPU}} = L_{\text{CPU}} + \frac{1}{T} \cdot\sum_{i = 1}^{\#cores} \sum_{j = 1}^{N_\text{CPU}} x_{ij}\cdot w_{ij}
\end{aligned}
\end{equation*}

\section{Evaluation}
\label{sec:experiments}

\glsreset{mape}

In this section, we evaluate the holistic power modeling approach discussed in \Cref{sec:model}, and its in-kernel implementation within the \runmeter online monitoring framework described in \Cref{sec:approach}.

\subsection{Experimental methodology}
\label{subsec:exp-method}

The target platform for our experiments is an \xavierboard, powered by the Xavier \gls{soc}~\cite{nvidiaxavier}. It is a highly parallel and heterogeneous \gls{soc} provided with an 8-core 64-bit ARMv8.2 CPU, a 512-core NVIDIA Volta \gls{gpu}, and several additional accelerators for deep-learning, computer vision, and video encoding/decoding. With many \gls{dvfs} \emph{power profiles} available for its sub-systems, this platform represents a challenging state-of-the-art target to explore our approach. In particular, the single CPU island on the platform can be clocked at 29 different discrete frequencies between \SI{115}{\mhz} and \SI{2.3}{\giga\hertz}, while the \gls{gpu} has 14 available \gls{dvfs} states between \SI{115}{\mhz} and \SI{1.4}{\giga\hertz}.

For the evaluation of our power modeling approach, we target the CPU and \gls{gpu} sub-systems,
$$D^\ast = \{CPU,\ GPU\}$$
considering the following \gls{dvfs} states:
\begin{gather*}
    F_{\text{CPU}}^\ast = \{ \SI{730}{\mega\hertz},\; \SI{1.2}{\giga\hertz},\; \SI{2.3}{\giga\hertz} \}
    \\
    F_{GPU}^\ast = F_{GPU} = \{ \text{all 14 from \SI{115}{\mega\hertz} to \SI{1.4}{\giga\hertz}} \}
\end{gather*}
To build the input dataset, we profile several workloads on the target platform based on the considerations of \Cref{subsec:model-benchmarks}. For the CPU, we employ 17 different OpenMP benchmarks from the Rodinia 3.1 heterogeneous benchmark suite~\cite{che2009rodinia} in several multi-thread configurations and five additional synthetic benchmarks. For the \gls{gpu}, we employ ten different CUDA benchmarks from Rodinia. To average out possible interference in our measurements, such as unpredictable \gls{os} activity on the CPU, each workload is profiled 3 times.

\gls{pmc} samples are acquired in a continuous, periodical mode with a sampling period of \SI{100}{\milli\second}. During each sampling period,  power measures of the CPU and GPU sub-systems are also acquired from the INA3221 built-in power monitors~\cite{ina3221}. This grants the time correlation needed for an effective correlation analysis and training~\cite{malony2011parallel}. We find that collecting more than one sample per \SI{100}{\milli\second} does not capture any additional information due to the electrical inertia of the built-in current sensors.

As mentioned in \Cref{sec:background}, built-in power monitors are typically not robust tools for online, power-aware actuation policies. This is mainly due to their speed, coarse granularity, and low resolution, which for the Xavier is limited to about \SI{200}{\milli\watt}.
However, they are helpful for building datasets to achieve higher introspection, time granularity, and responsiveness enabled by \gls{pmc}-based power models, as proved in \Cref{sec:experiments}.

\subsection{Offline Platform Characterization and Modeling}
\label{sec:results-characterization}

This section discusses the result of the platform characterization and power modeling (\Cref{sec:model}) of the individual CPU and GPU sub-systems for the NVIDIA Jetson AGX Xavier case study.
In the interest of conciseness, in the following, we only analyze the main results to support our subsequent
discussion. A thorough report is available in \cite{Mazzola2022SAMOS}.

\subsubsection{Sub-system Characterization}
\label{subsec:platform_char}

For the CPU sub-system, the results of the platform characterization suggest that the power consumption of the cores is highly correlated, depending on the selected \gls{dvfs} state, with the number of cycles of activity, the number of retired instructions, the floating point activity, and various cache-related events. From our experiments, the counters that correlate best with the measured power consumption results are all cross-compatible. Therefore, for the power model parameter selection, we consider the three best counters for each frequency as the maximum allowed by the ARM \gls{pmu}. In addition, the ARM \gls{pmu} always exposes a CPU cycle counter, which we include in the final power model.

For the \gls{gpu} sub-system of the Xavier \gls{soc}, our results expose multiple incompatibilities among the counters that would best correlate with the power profile. To be able to simultaneously track the optimal model parameters at runtime, we employ the \gls{pmu}-aware algorithm described in \Cref{sec:model-characterization-steps}, which automatically selects the subset of cross-compatible counters that best correlate with the power consumption.
We find that counters related to L2 cache utilization and warp execution best correlate with the power consumption of the \gls{gpu}. By enhancing the counter selection with feedback from the GPU model validation, we find that eight \glspl{pmc} per frequency is the optimal trade-off between the number of independent variables, affecting the model evaluation time and the model accuracy.

\subsubsection{Sub-system Modeling and Validation}
\label{subsec:train_valid}

\begin{figure*}[th!]
    \centering
    \includegraphics{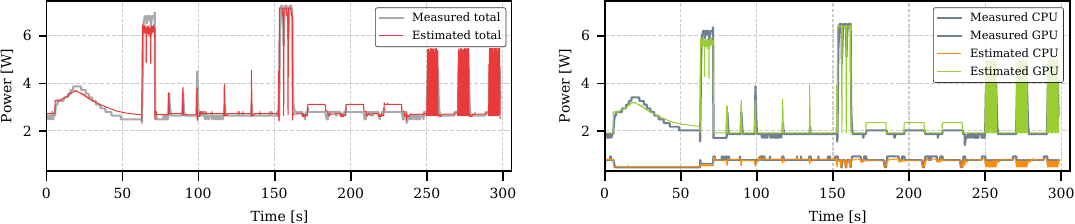}
    \caption{Instantaneous power estimate for the system-level power model (on the left) and its breakdown to the individual sub-systems (on the right), with $f_{\text{CPU}}=\SI{1.2}{\giga\hertz}$, $f_{\text{GPU}}=\SI{830}{\mhz}$.}
    \label{fig:comb_model_power}
\end{figure*}

For the CPU, we train a linear model based on \Cref{eq:lut} individually for each frequency, with a \gls{nnls} regression. We employ four independent variables per core, i.e., the three configurable \glspl{pmc} for each frequency and the cycle counter. Out of our input dataset, we use a random selection of \SI{70}{\%} of the total data for training and the remaining \SI{30}{\%} for validation. In terms of instantaneous power accuracy, the model achieves a \gls{mape} between \SI{3}{\%} and \SI{4.4}{\%} based on the frequency, with a standard deviation of approximately \SI{5}{\%}. When employed to estimate the energy over the full validation set, our model achieves a maximum error of \SI{4}{\%}, delivering an equal or superior accuracy with respect to the state of the art.

For the \gls{gpu}, we likewise train the \Cref{eq:lut} for each of the 14 \gls{gpu} frequencies with a \gls{nnls} linear regression. We use the same 70\% and 30\% ratio for the training and validation set. Comparing the instantaneous power consumption estimation with the data measured on the real platform, we obtain a \gls{mape} between \SI{6}{\%} and \SI{8}{\%}, depending on the frequency. The standard deviation over all frequencies is approximately \SI{8}{\%}. The maximum energy estimation error over the full validation set is \SI{5.5}{\%} over all frequencies, with an average of \SI{2.2}{\%}.

\subsection{Combined Model Evaluation}
\label{subsec:comb_model_res}

After building, training, and validating the CPU and \gls{gpu} power models individually, we combine them to obtain a system-level power model for every possible combination of $f_{\text{CPU}} \in F_{\text{CPU}}^\ast$ and $f_{\text{GPU}} \in F_{\text{GPU}}^\ast$, corresponding to the LUT of \Cref{eq:lut}.

\Cref{fig:comb_model_power} shows how our decomposable power model can effectively track the instantaneous power consumption of the system over time. The achieved instantaneous power \gls{mape} of the final, combined model has an average of \SI{8.6}{\%} over all CPU and \gls{gpu} frequency combinations.
Regarding energy, the model reaches an average estimation error of \SI{2.5}{\%}.
However, our results highlight that the energy estimation error of the combined model is higher when $f_{CPU}$ and $f_{GPU}$ diverge from each other. In particular, when $f_{GPU}$ is very low compared to $f_{CPU}$, the CPU may stall waiting for the offloaded computation. While our power model does not capture this behavior, a real scenario where this occurs is highly unlikely due to its inefficiency.

Therefore, by considering all CPU and GPU frequencies combinations such that  $f_{\text{GPU}}>\SI{600}{\mhz}$, we report an instantaneous power \gls{mape} of \SI{7.5}{\%} and energy estimation error of \SI{1.3}{\%}, with a maximum of \SI{3.1}{\%}.

\subsection{CPU Power Monitoring with Runmeter}
\label{sec:results-runmeter}

This section discusses the evaluation of our online monitoring framework, \runmeter.
To evaluate \runmeter, as a case study, we integrate it in the kernel of the Linux distribution running on the NVIDIA Jetson AGX Xavier. Then, through \runmeter, we implement the power monitor discussed in \Cref{sec:approach} with support for the CPU sub-system.

First, we discuss the impact of porting models such as the one derived in \Cref{subsec:cpu_model} from floating-point to fixed-point arithmetic,  which is necessary for their in-kernel implementation. Then, we deploy the CPU sub-system model on the target platform and compare the online power estimation provided by \runmeter with the measures from the onboard analog sensors. Finally, we include an analysis of the overhead introduced to the system by \runmeter, proving that both our power modeling methodology and our monitoring framework are ideal for online policy actuation such as \gls{dpm} and power-aware task scheduling.

\subsubsection{Fixed-point Approximation Error}
\label{subsec:exp-fixp}

In this section, we evaluate the approximation error introduced by our fixed-point implementation of the power model described in \Cref{sec:approach}, necessary to integrate it as part of a kernel module. For the fixed-point implementation, we use 64-bit integers, assigning the $29$ less significant bits to the fractional part.

To analyze the approximation error, we implement the power model in user space as a generic C++ procedure, which can be applied to any numeric type.
From user space, we collect the data published by the \runmetermodule and feed them to the C++ power model, which evaluates it through the floating-point and the fixed-point power models to compute the approximation error. For this evaluation, we use the same validation set discussed in \Cref{sec:results-characterization}.

\Cref{fig:fixp-error} shows the distribution of the approximation error between the floating-point and the fixed-point implementation.
From our extensive evaluation, the maximum absolute approximation error is about \SI{17}{\milli\watt}; the mean error, however, is only of about \SI{0.17}{\milli\watt}. The maximum percentage error is always below \SI{0.8}{\%} of the power consumption estimated using floating-point arithmetic, with a mean error of about \SI{0.015}{\%}.
Given the negligible magnitude of the error introduced by the fixed-point implementation, we conclude that this approximation does not impact the accuracy of the model in any meaningful way.

\begin{figure}[t]
    \centering
    \includegraphics[]{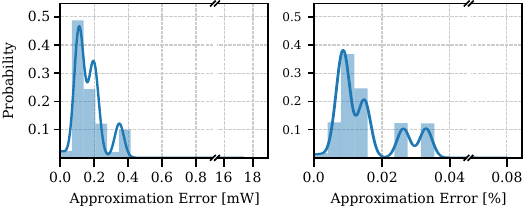}

    \caption{
        Distribution of approximation error between floating-point and fixed-point implementations of the CPU power model. The distribution is shown in terms of both absolute power approximation error and percentage error.
    }
    \label{fig:fixp-error}

\end{figure}

\subsubsection{Online Power Estimation Accuracy}
\label{subsec:exp-runmeter}

Employing the fixed-point implementation of the CPU power model analyzed in the previous section, we integrate the power monitor in the \runmeterframework.
We then log the data advertised at runtime by the power model in \runmeter to later perform post-mortem analysis.
Therefore, differently from the discussion so far, the power estimates analyzed in this section are computed directly at runtime as soon as new \gls{pmc} samples are available.

\Cref{fig:energy-error-distribution} shows the \gls{ape} distribution of the energy estimation provided by the in-kernel model when compared against the value collected from the onboard analog sensor for all benchmarks. In general, the maximum \gls{ape} registered over all our experiments is around \SI{29}{\%}, the error at \nth{90} percentile is around \SI{20.8}{\%}, and the \gls{mape} is around \SI{9}{\%}. The input dataset is the same as the validation set discussed in \Cref{sec:results-characterization}.

\begin{figure}[tp]
    \centering
    \includegraphics{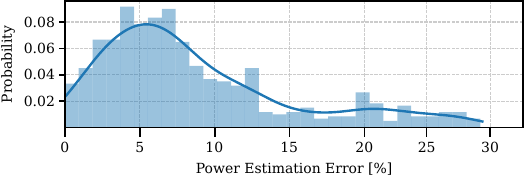}
    \caption{Distribution of the \gls{ape} of the online energy estimates over the duration of each benchmark.}
    \label{fig:energy-error-distribution}
\end{figure}

\begin{figure}[t]
    \centering
    \subcaptionbox{\SI{730}{\mega\hertz}\label{fig:cpu_online_730}}{
        \includegraphics[]{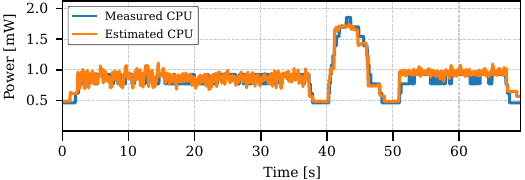}
    }

    \vspace*{.5em}

    \subcaptionbox{\SI{1.2}{\giga\hertz}\label{fig:cpu_online_1200}}{
        \includegraphics[]{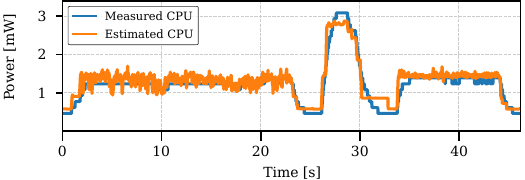}
    }

    \vspace*{.5em}

    \subcaptionbox{\SI{2.3}{\giga\hertz}\label{fig:cpu_online_2300}}{
        \includegraphics[]{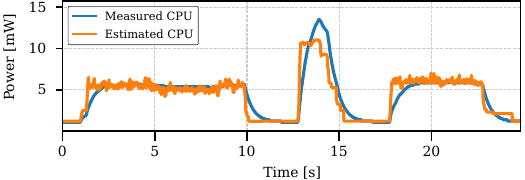}
    }

    \caption{
        Comparison of the instantaneous CPU power consumption measurement provided by the onboard INA3221 sensor and the estimation computed at runtime by the in-kernel power model. Each plot represents the same sequential execution of several workloads over time at different frequencies.
    }
    \label{fig:energy-estimation}
\end{figure}

It must be noted, however, that most of the estimation error accounted for during such an evaluation can be attributed to very specific time frames when the phase of the workload abruptly changes.
Such behavior is visible in the example CPU power profiles depicted in \Cref{fig:energy-estimation}.
On sharp changes of the system activity, corresponding to rapid switches of the power consumption, the power estimated by the \gls{pmc}-based power model has faster rising and falling edges than the power measured by the analog sensor. This is especially visible at higher CPU frequencies, where the inertia of the analog current sensors has an increasingly higher impact on the measurements due to the faster CPU activity.
On the other hand, \glspl{pmc} are embedded in the digital domain, and their values instantly reflect the dynamic behavior of the profiled workloads.
Nevertheless, our power modeling approach uses the onboard power sensor to build the input dataset for training and validation. While this makes the procedure automatic, easier, and less error-prone, the error on the estimation provided by our model over time is unavoidably high during transients in which the power sensor is too slow to react. As a matter of fact, the occasionally high estimation error is strictly limited to these transient conditions, where the high responsiveness of our \gls{pmc}-based model can accurately track even brief fluctuations in the power profile.

Therefore, the problem emerges of how to make sure that the power values estimated during transient by our \gls{pmc}-based model are actually reliable, if they have been trained with the INA3221 measurements themselves. As a solution, during the training phase, we deliberately bias the training set towards workloads with more stable activity: this means that the power model is trained, on average, with power values matching the actual consumption of the platform. This decouples the trained weights of the model from the low sensibility and time granularity of the input power data used for training. Once trained, the power model can scale and interpolate those values accordingly to the \gls{pmc} samples collected at runtime, providing faster power estimation and higher responsiveness.

\subsubsection{Monitoring Overhead}
\label{subsec:exp-overhead}

Being integrated with callbacks triggered at specific times during the Linux kernel execution, the \runmeterframework imposes a certain processing overhead mainly due to \gls{pmc} data collection and manipulation, including model estimation.

To measure this overhead, we profile the execution of the \runmetermodule callbacks, which are the only additional components with respect to a vanilla Linux kernel. We perform these measurements in various working conditions, ranging from an \quotes{idle} state to the execution of multiple parallel applications from the set of benchmarks described in
\cite{Mazzola2022SAMOS}. We use the same frequencies employed for the CPU model evaluation.
The maximum overhead is reported when many applications execute concurrently on the system, as the number of invocations of \runmeter's callbacks increases with the number of context switches performed by the system.
In the worst-case condition of intense context switching, the time spent executing all of the framework's callbacks never exceeds \SI{7}{\milli\second} per second (i.e., \SI{0.7}{\%} overhead). Moreover, the execution of all framework's callbacks significantly speeds up when increasing the CPU frequency, reducing to less than \SI{2}{\milli\second} per second in the worst case (\SI{0.2}{\%} overhead) when operating at \SI{2.3}{\giga\hertz}.
In idle conditions, the overhead of the framework at \SI{2.3}{\giga\hertz} is always less than \SI{0.4}{\milli\second} per second (\SI{0.04}{\%}).

\section{Conclusions and Future Work}
\label{sec:conclusions}

With this work, we propose a systematic, data-driven, and architecture-agnostic approach to \gls{dvfs}-aware statistical power modeling of heterogeneous computing systems. Our approach individually models each sub-system through its local \glspl{pmc}, autonomously selecting the best ones to represent its power consumption. The sub-system models are later re-composed in a LUT-based system-level power model, able to grasp the complex behaviors of \gls{dvfs}-enabled hardware with simple, linear models. This approach achieves an unprecedented combination of general applicability, automated model construction, lightweight model evaluation, and high accuracy.

In addition, we propose \runmeter, a novel integrated framework for the evaluation of the generated models online from within the Linux kernel. \runmeter is a substantial improvement over existing mechanisms based on \gls{pmc} tracking as it focused on minimizing the response time between \gls{pmc} observation and the evaluation of the model, without introducing any substantial overhead.

The validation of our power modeling approach on the state-of-the-art \xavierboard embedded platform results in power and energy estimation accuracies aligned or superior with respect to state-of-the-art works. %
Additionally, the model shows desirable properties of responsiveness and decomposability. By integrating the \runmeter framework in the Linux kernel of the same platform, we also prove the viability of our modeling and monitoring approach for online power tracking, a key prerequisite for implementing power-aware control loops in \gls{dpm} and power-aware task scheduling.

These results pave the way for further work to bring the benefits of this modeling approach to additional components in the Linux kernel, through the \runmeter framework. In particular, we plan to investigate the possible integration of our monitoring framework with the Linux real-time task scheduler (\ccode{SCHED\_DEADLINE}), together with the CPU frequency governor, with the aim to improve the effectiveness and correctness of energy-aware real-time task scheduling within Linux.

Further directions of work also include going beyond the estimation of the current hardware status through predictive models. As of now, the Linux kernel contains very simple linear models for estimating the power consumed at each frequency, used to make decisions when selecting the appropriate frequency at which the CPU should execute.
Models based on online \gls{pmc} data, like that collected by the \runmeter framework, may prove to be more effective from an energy-saving perspective while maintaining a very low overhead.

\ifACM{}
\ifIEEE{}

%

\newpage
\begin{IEEEbiography}[{\includegraphics[width=1in,height=1.25in,clip]{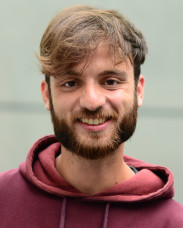}}]{Sergio Mazzola}
received his M.Sc. degree in Electrical Engineering and Information Technology at the Polytechnic University of Turin (Italy) in 2021. Currently, he is pursuing a Ph.D. degree with the Digital Circuits and Systems group of Luca Benini at ETH Zürich (Switzerland). His research interests include high-performance computer architecture, focusing on massively parallel systems and low-power design.
\end{IEEEbiography}

\vspace{-8mm}
\begin{IEEEbiography}[{\includegraphics[width=1in,height=1.25in,trim=7.5cm 12.5cm 7.5cm 5cm,clip,keepaspectratio]{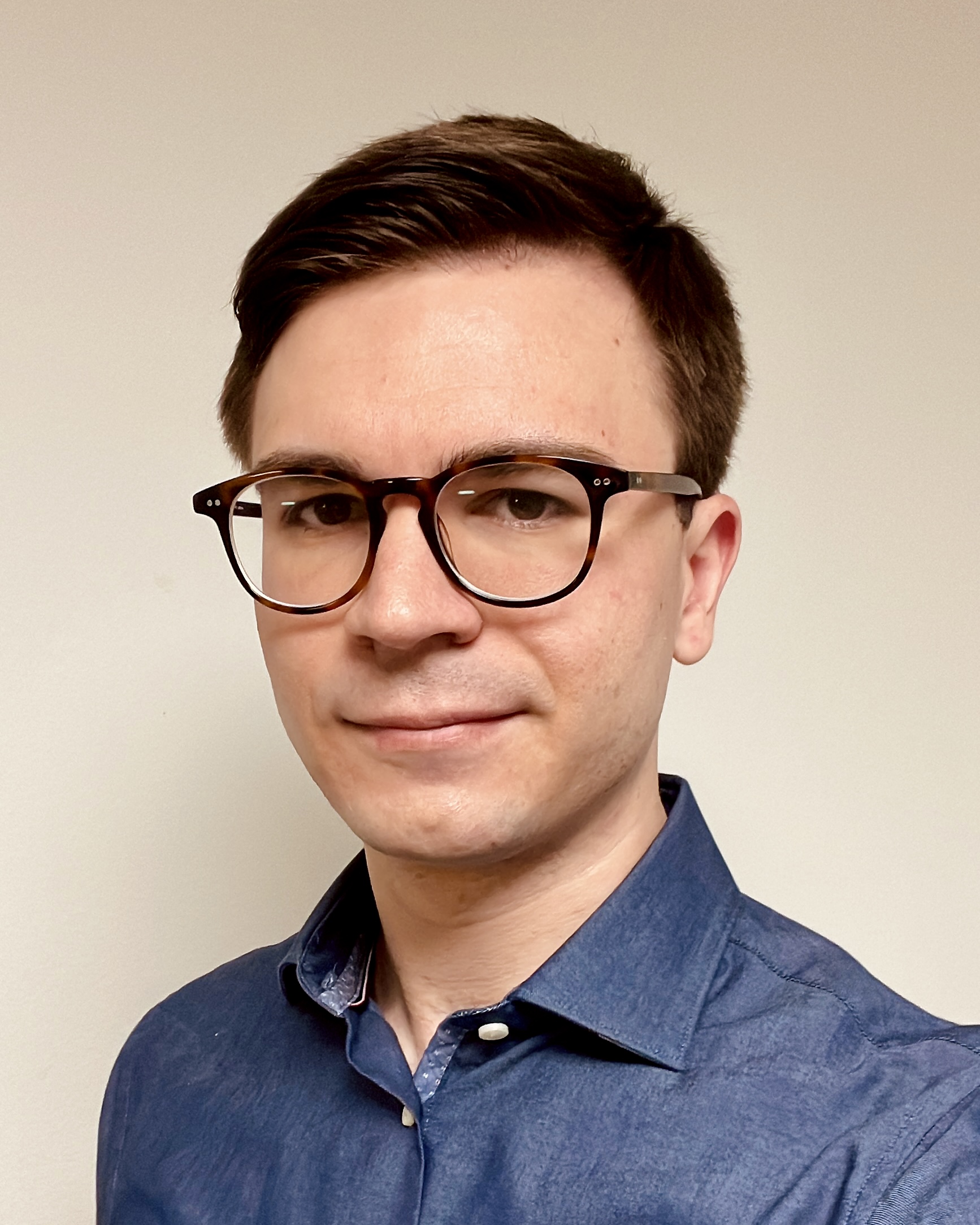}}]{Gabriele Ara}
  received his M.Sc. degree in Embedded Computing Systems from the University of Pisa and the Scuola Superiore Sant'Anna (Italy). He received his Ph.D. in Computer Engineering from the Scuola Superiore Sant'Anna, where he investigated novel OS mechanisms for energy-efficient real-time and high-performance networking applications. Since 2023 he is a Postdoctoral Researcher at SSSA in the Real-Time Systems Lab, where he studies techniques to estimate and mitigate the energy consumption of real-time tasks on embedded platforms.
\end{IEEEbiography}

\vspace{-8mm}
\begin{IEEEbiography}[{\includegraphics[width=1in,height=1.25in,clip]{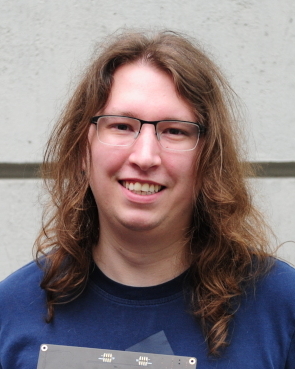}}]{Thomas Benz}
  (Graduate Student Member, IEEE) received his B.Sc. and M.Sc. degrees in Electrical Engineering and Information Technology from ETH Zürich in 2018 and 2020, respectively. %
  He is currently pursuing a Ph.D. degree in the Digital Circuits and Systems group of Prof.\ Benini. %
  His research interests include energy-efficient, high-performance computer architectures and the design of ASICs.
\end{IEEEbiography}

\vspace{-8mm}
\begin{IEEEbiography}[{\includegraphics[width=1in,height=1.25in,clip]{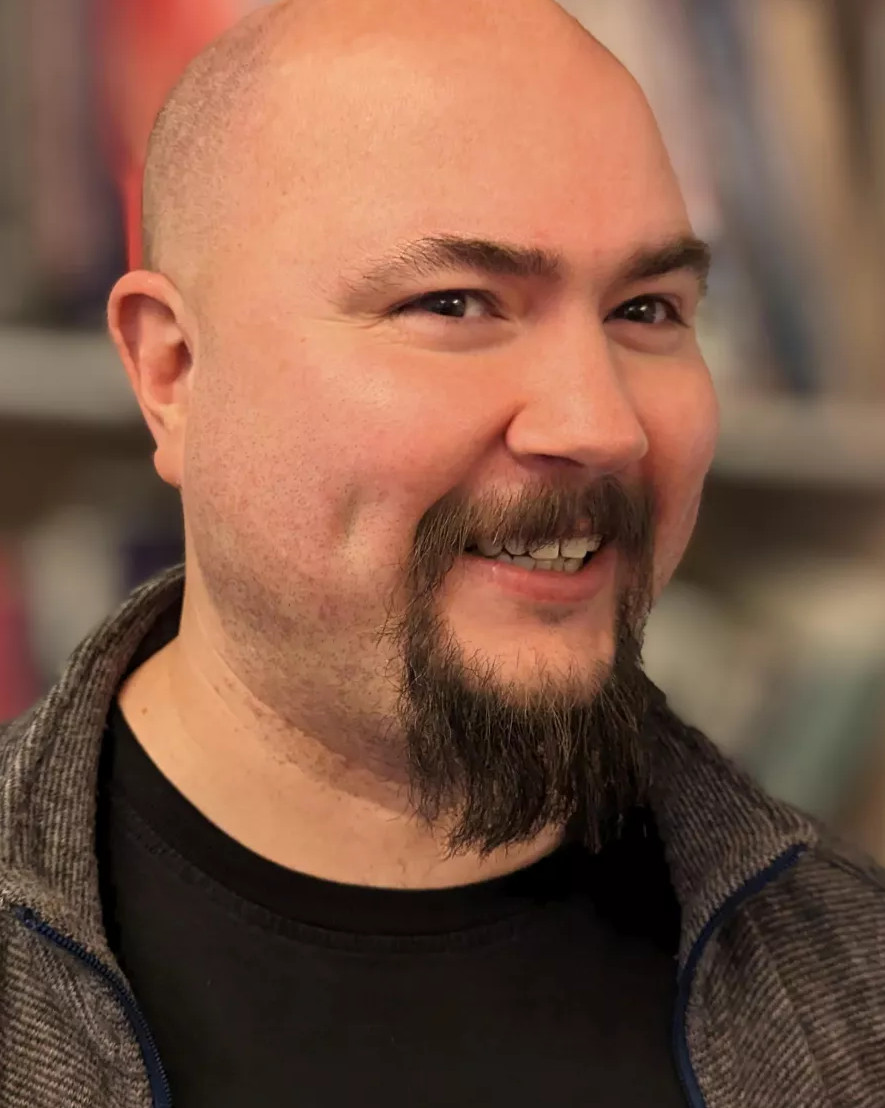}}]{Bj{\"o}rn~Forsberg} received his M.Sc. degree from Uppsala Universitet (Sweden) in 2015, and his Ph.D. degree from ETH Zürich in 2021, for his work on timing-predictable execution for heterogeneous embedded real-time systems based on compiler and software-centric techniques for real-time guarantees on modern high-end embedded hardware. His interests lie in new and enabling technology at the hardware and software boundary, and its impact
on programmability, real-time, energy-efficiency, and performance.
\end{IEEEbiography}

\vspace{-8mm}
\begin{IEEEbiography}[{\includegraphics[width=1in,height=1.25in,clip]{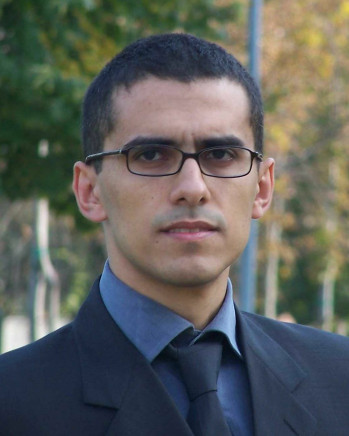}}]{Tommaso Cucinotta}
(Member, IEEE) received his M.Sc. degree in Computer Engineering from the University of Pisa and his Ph.D. degree in Computer Engineering from Scuola Superiore Sant’Anna, where he has been investigating real-time scheduling for soft real-time and multimedia applications, and predictability in infrastructures for cloud computing and NFV. He has been MTS in Bell Labs in Dublin (Ireland), investigating security and real-time performance of cloud services. He has also been a software engineer at Amazon Web Services in Dublin, where he worked on improving the performance and scalability of DynamoDB. Since 2016, he is an Associate Professor with Scuola Superiore Sant’Anna and head of the Real-Time Systems Lab (RETIS) since 2019.
\end{IEEEbiography}

\vspace{-8mm}
\begin{IEEEbiography}[{\includegraphics[width=1in,height=1.25in,clip]{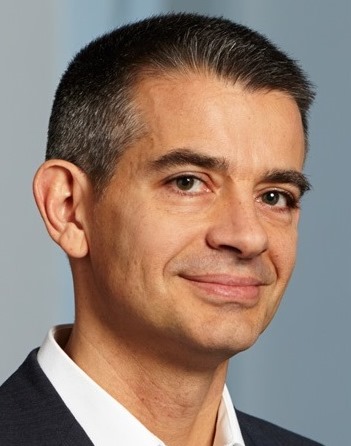}}]{Luca Benini}
(Fellow, IEEE) is the Chair of Digital Circuits and Systems at  ETH Zürich and a Full Professor at the University of Bologna.
He has served as Chief Architect for the Platform2012 in STMicroelectronics, Grenoble.
Dr.\ Benini's research interests are in energy-efficient systems and multi-core SoC design.
He is also active in the area of energy-efficient smart sensors and sensor networks.
He has published more than \num{1000} papers, five books and several book chapters.
He is a Fellow of the ACM and a member of the Academia Europaea.
\end{IEEEbiography}

\vfill

\end{document}